\documentclass[12pt,reqno]{amsart}
\usepackage{amsmath,amssymb,amsfonts,amsthm}
\usepackage[mathscr]{eucal}
\usepackage[russian,english]{babel}
\usepackage[all]{xy}
\usepackage{hyperref}
\usepackage{cite}
\usepackage{ulem,soul}
\usepackage{cmap}

\usepackage{color}

\title[Classical and Quantum Stability of higher-derivative dynamics]
{Classical and Quantum Stability of higher-derivative dynamics}
\author{D.S. Kaparulin, S.L. Lyakhovich and  A.A. Sharapov}
\address{Physics Faculty, Tomsk State University, Tomsk 634050, Russia}
\email{dsc@phys.tsu.ru, sll@phys.tsu.ru, sharapov@phys.tsu.ru}

\begin{document}


\begin{abstract}
We observe that a wide class of higher-derivative systems admits a
bounded integral of motion that ensures the classical stability of
dynamics, while the canonical energy is unbounded. We use the
concept of a Lagrange anchor to demonstrate that the bounded
integral of motion is connected with the time-translation
invariance. A procedure is suggested for switching on interactions
in free higher-derivative systems without breaking their stability.
We also demonstrate the quantization technique that keeps the
higher-derivative dynamics stable at quantum level. The general
construction is illustrated by the examples of the Pais-Uhlenbeck
oscillator, higher-derivative scalar field model, and the Podolsky
electrodynamics. For all these models, the positive integrals of
motion are explicitly constructed and the interactions are included
such that keep the system stable.
\end{abstract}

\maketitle

\section{Introduction} The higher-derivative dynamics are as good
as the conventional ones in many principal issues. In particular,
the Noether theorem still applies that connects symmetries and
conservation laws. The Hamiltonian formulation is also known for
both non-singular theories \cite{Ost} and the most general
higher-derivative Lagrangians with singular Hessian \cite{GLT}. For
many decades, a variety of higher-derivative models are studied once
and again. The long known examples include the Pais-Uhlenbeck
oscillator \cite{PU}, Podolsky electrodynamics \cite{Podolsky1942,
Podolsky1944, PS}, various conformal field theories
\cite{Weyl,FradkinTseytlin1985}, $R^2$-gravity
\cite{R2,Starobinsky1980}, and many others. The vast literature
exists on various higher derivative models, we can mention the
papers \cite{Smilga2005, SMILGA2007, BENDER2007, BENDER2008,
SMILGA2014, BOLONEK2005,DAMASKINSKY2006, Mostafazadeh2010,
Ketov2011, PAVSIC2013, PAVSIC2013b, AEBV, Bufalo2011, Bufalo2012,
Bufalo2013, MCHS, Vasiliev2009, Francia2012, Joung2012, BKRTY2011,
JM2012, METSAEV2014, BL, AStrominger, Deser2013, BHT2010,
Faraoni2005, fREnergy2, fRGravityRev1, fRGravityRev2,
Plyushchay1989, Chen2013, Andrzejewski2007} and references therein.

In many cases, the higher derivative models reveal remarkable
properties. They often admit a wider symmetry than the
first-derivative analogues. One more typical phenomenon is that the
inclusion of the higher derivatives in Lagrangian can improve the
convergency in field theoretical models both at classical and
quantum level.

A notorious difficulty of higher derivative models concerns
instability of their dynamics. The Noether energy is typically
unbounded for higher derivative Lagrangians, and this fact is
usually considered as an evidence of classical instability. At
quantum level, the instability reveals itself by ghost poles in the
propagator and related problem with the unbounded spectrum of
energy. In their turn, the problems of quantum instability are
related to the fact that Ostrogradsky's Hamiltonian, being the phase
space equivalent of Noether's energy, is unbounded due to the higher
derivatives.

For the general acceleration dependent Lagrangian, the Noether
energy
\begin{equation}\label{NE}
     E_{\mathfrak{N}}\equiv\left(\frac{\partial L}{\partial \dot{\phi}^i} -\frac{d}{dt}\frac{\partial L}{\partial
    \ddot{\phi}^i} \right) \dot{\phi}^i + \frac{\partial L}{\partial \ddot{\phi}^i}
       \ddot{\phi}^i  - L
\end{equation}
cannot be positive because of a simple reason: it is linear in
$\dddot{\phi}{}^i$. The third derivatives are the independent
initial data for the fourth order Lagrange equations whenever the
Hessian $$\frac{\partial^2 L(\phi, \dot{\phi},
\ddot{\phi})}{\partial\ddot{\phi}^i\partial \ddot{\phi}^j}$$ is
nondegenerate.

For the models with degenerate Hessian, the constraints appear in
phase space \cite{GLT}, that can restrict the third derivatives. It
is a very special case, where the constraints are strong enough to
make the linear function positive, though it may happen on some
occasions \cite{Chen2013, Andrzejewski2007}. The known examples of
this type include the higher-order theories of gravity
\cite{Deser2013,AStrominger, Faraoni2005,fREnergy2} and some models
of higher spin fields \cite{BHT2010,BKRTY2011,JM2012}. One more
example is given by the relativistic point particle, whose
Lagrangian linearly depends on the curvature of world line
\cite{Plyushchay1989}. Because of positive Hamiltonian, these models
are stable classically and have no ghosts at the quantum level.

The positivity of canonical Noether's energy is a sufficient
condition for classical stability, while it is unnecessary. The
simplest example is provided by the Pais-Uhlenbeck oscillator. The
Lagrangian is acceleration-dependent and non-singular. Therefore
Noether's energy is unbounded in this model, while the classical
stability is obvious, because the motion is bounded. The point is
that the Pais-Uhlenbeck oscillator admits another integral of motion
which is positive. It is the integral which provides stability.
Various specific reasons can be seen for considering this positive
conserved quantity as a natural candidate for the role of energy in
this model. We elaborate on the details in the next section.

In this paper, we consider the issue of stability of the
higher-derivative theories from the viewpoint of existence of a
positive integral of motion. In the first instance, we consider a
class of linear higher-derivative systems. The fourth-order operator
of the equations is supposed to admit \textit{factorization} into a
pair of different second-order operators satisfying certain (not too
restrictive) condition. Many of known higher-derivative linear
models fall into this class, including the Pais-Uhlenbeck
oscillator, Podolsky electrodynamics, linearized conformal gravity.
For the models of this type we construct the integral of motion
which is squared in third derivatives. It can be either bounded or
unbounded depending on signature, in contrast to the Noether energy,
which is almost always unbounded unless the theory is not strongly
constrained. Besides the general method of construction, we
explicitly present the positive integral in several
higher-derivative models with unbounded Noether's energy. As we
further demonstrate, the concept of factorization extends beyond the
linear level providing the procedure for inclusion of stable
interactions in higher-derivative theories.

As the next step we establish a relationship between the conserved
positive quantity, being responsible for the classical stability of
the higher-derivative dynamics, and the translation invariance. The
key tool allowing one to connect the integral of motion with the
symmetry is the concept of a \textit{Lagrange anchor} \cite{KazLS}.
Originally, the Lagrange anchor\footnote{To make the article
self-contained, we provide some generalities on the Lagrange anchor
in Appendix \ref{AppA}.} was introduced as a tool for extending the
BV-BRST quantization procedure beyond the scope of Lagrangian
theories \cite{KazLS}. Given not necessarily variational equations
of motion, the Lagrange anchor allows one to define the
Schwinger-Dyson equation \cite{LS1} and the path integral
representation for the partition function \cite{LS2}. It has been
later noticed that the Lagrange anchor maps conservation laws to
symmetries \cite{KapLS2010} extending in such a way the Noether
theorem beyond the class of variational equations. Any Lagrangian
system admits a canonical Lagrange anchor, which is is given by an
identity operator. The same system of equations may admit different
inequivalent Lagrange anchors. Inequivalent Lagrange anchors result
in inequivalent quantum theories, and different Lagrange anchors
assign different symmetries to the same conservation law. It turns
out that the higher-derivative Lagrnagian dynamics of the considered
class always admit the Lagrange anchor which is inequivalent to the
canonical one. If the energy is connected to the time-translation
invariance with this anchor, we arrive at positive energy which
differs from the unbounded expression (\ref{NE}). Furthermore, the
quantization with this anchor will not break the stability as we
explain below.

For the first-order unconstrained mechanical systems without gauge
symmetries, each Lagrange anchor defines and is defined by a
bi-vector \cite{KazLS, KapLS2013, BG2011a}. This means, in
particular, that when a non-singular, higher-derivative Lagrangian
of a mechanical system\footnote{For the gauge invariant and/or
constrained mechanical systems, the connection between Lagrange
anchor and Poisson structures is more involved. A Lagrange anchor in
this case gives rise to a weak Poisson structure \cite{LS0}. In the
field theory, the relationship is even more complex and it is not
completely known at the moment. } is reduced to the first order by
introducing auxiliary variables, the first-order system will be
bi-Hamiltonian whenever the two inequivalent Lagrange anchors are
admissible for the higher-derivative equations. The different
Hamiltonians represent in the phase space the different conserved
quantities connected with the time-shift transformation by different
Lagrange anchors in the configuration space. The fact that the
Pais-Uhlenbeck oscillator is a bi-Hamiltonian system has been
noticed in \cite{BOLONEK2005,DAMASKINSKY2006}. The
``non-Ostrogradsky Hamiltonian'' is positive. As we observe, it
corresponds to the integral of motion connected with the time-shift
symmetry of the Pais-Uhlenbeck oscillator by an alternative Lagrange
anchor. As we will demonstrate, it is not an isolated observation
which is valid for particular higher-derivative model. It is a part
of a broader picture concerning the issue of stability in the
higher-derivative systems. These systems turn out to be classically
stable because of the same reason as the first-derivative Lagrangian
dynamics: they all have a positive energy that conserves. The only
essential difference is that the definition of energy may involve a
more general Lagrange anchor than the canonical one.

In this paper, we also address the problem of including interaction
without breaking stability of higher derivative dynamics. For the
Lagrangian equations without higher derivatives, and with a positive
Noether's energy, it would be sufficient to include the
translation-invariant interaction into the Lagrangian in  a way that
keeps the energy bounded. For the general higher-derivative systems,
where stability cannot be controlled by Noether's energy (\ref{NE})
anymore, the issue becomes more tricky. As we see, a positive
(non-canonical) energy is connected with the translation invariance
by a non-canonical Lagrange anchor in the higher-derivative theory.
With this regard, the sufficient conditions for stability mean to
meet the following requirements, which are automatically satisfied
with the canonical anchor. First, the interaction has to be included
simultaneously into the equations of motion and in the Lagrange
anchor to keep them compatible. When a relevant Lagrange anchor is
canonical, it is automatically compatible with the Lagrangian
vertices in the equations. For the stability of higher-derivative
systems, as we see, typically a non-canonical Lagrange anchor is
relevant because it connects the positive integral of motion with
translation invariance. 
Second, the
interaction should keep the positivity of the energy. If the vertex
is Lagrangian and translation invariant, this will mean that the
Noether energy still conserves, though it does not automatically
mean the same for a positive energy which is a different integral of
motion. The requirement for the deformed energy to conserve and keep
being positive is an additional requirement imposed on the
interaction. The last but not least, the deformed Lagrange anchor
should connect the positive energy of interacting system with the
generator of time translations. This is not automatically satisfied
either. We demonstrate by examples that all these requirements can
be met, though the stability control is not so simple procedure as
it is in the theories without higher derivatives.

The paper is organized as follows. In the next warming-up section we
consider the model of the Pais-Uhlenbeck oscillator to illustrate
the key general constructs we further use to control the stability
of higher-derivative dynamics. Section 3 describes the general
structure of the factorable higher-derivative dynamics, both linear
and non-linear, that allows one to control stability at the
classical level and keep it upon quantization. Section 4 illustrates
the proposed technique by examples of higher-derivative scalar field
model and Podolsky's electrodynamics. We demonstrate stability of
these models. As the paper essentially employs the Lagrange anchor
method developed in \cite{KazLS,LS1,LS2,KapLS2010,KapLS2013}, we
outline the relevant aspects of this construction in the appendices,
to make the paper self-contained. The general idea of a Lagrange
anchor is explained in Appendix \ref{AppA}. This Appendix also
provides some relations, which are used in this work. Appendix B
demonstrates how the Lagrange anchor is applied to connect conserved
quantities with symmetries. A particular consideration is given to
the possibility to connect different conserved quantities to the
translation invariance when the system admits different anchors.
Appendix C provides an elementary technique of finding the Lagrange
anchors for free field equations. It also explains why the
higher-derivative dynamics admit a wider set of Lagrange anchors
than the second-order field equations. Appendix D explains how the
linear techniques for finding the Lagrange anchors are extended to a
certain class of non-linear higher-derivative systems considered in
this paper. The Appendices provide the background and techniques for
those who wish to apply or further develop the method, while the
results of the present paper can be apprehended by consulting only
the relations which are directly referred to in the main text.

\section{Stability of the Pais-Uhlenbeck oscillator}\label{1}

In this  section, we consider the Pais-Uhlenbeck (PU) oscillator
which is studied for decades, see \cite{Smilga2005, SMILGA2007,
BENDER2007, BENDER2008, SMILGA2014, BOLONEK2005,DAMASKINSKY2006,
Mostafazadeh2010, Ketov2011, PAVSIC2013, PAVSIC2013b} and references
therein. By this simplest model we exemplify the key structures
related to the (in)stability problem of higher derivative dynamics.
In the next section these structures are described in the general
form.

The action of the PU oscillator involves derivatives of a single
variable $\phi(t)$ up to the second order:
\begin{equation}\label{PUL}
S[\phi]= \int dt L\,, \qquad L=\frac{1}{2(\omega_1^2-\omega_2^2)}
\Big(\ddot{\phi}+\omega_1^2\phi\Big)\Big(\ddot{\phi}+\omega_2^2\phi\Big)\,;
\end{equation}
here $\omega_1\neq\omega_2$ are the frequencies of oscillations. The
corresponding equation of motion reads
\begin{equation}\label{LEPU}
    \frac{\delta S}{\delta \phi}\equiv\frac{1}{\omega_1^2-\omega_2^2}\left(\frac{d^2}{dt^2}
    +\omega_1^2\right)\left(\frac{d^2}{dt^2}+\omega_2^2\right) \phi=0\, .
\end{equation}
As is seen, the fourth-order operator of the equation factorizes
into the product of the second-order commuting operators. Because of
this factorization, the general solution to equation (\ref{LEPU}) is
given by the sum
\begin{equation}\label{phi=xi+eta}
    \phi=\xi+\eta\,,
\end{equation}
where the functions $\xi$ and $\eta$ satisfy the second-order
equations
\begin{equation}\label{PUxieta}
    \Big(\frac{d^2}{dt^2}+\omega^2_1\Big)\xi=0,\qquad
    \Big(\frac{d^2}{dt^2}+\omega^2_2\Big)\eta=0\, .
\end{equation}
Conversely, if $\phi$ is a solution to the original fourth-order
equation (\ref{LEPU}), then the expressions
\begin{equation}\label{xieta}
    \xi = \frac{\ddot{\phi}+\omega_2^2\phi}{\omega_2^2-\omega_1^2}\, ,
    \qquad \eta =\frac{\ddot{\phi}+\omega_1^2\phi}{\omega_1^2-\omega_2^2}\,
\end{equation}
obey the second-order equations (\ref{PUxieta}). Relations
(\ref{phi=xi+eta}) and (\ref{xieta}) establish a one-to-one
correspondence between the solutions to the fourth-order equation
(\ref{LEPU}) and the second-order system (\ref{PUxieta}).

The general solution for $\phi$ is a linear combination of the two
independent harmonic oscillations
\begin{equation}\label{osc}
\xi=A_1\sin{\omega_1 (t-t_1)}\,,\qquad \eta=A_2\sin{\omega_2
(t-t_2)}\,.\end{equation} Taking the linear combination of the
energies of the oscillations, we get a two-parameter family of
integrals of motion for the PU model
\begin{equation}\label{EPxieta}
    E_{\alpha,\beta}=\frac{\alpha}{2}\Big(\dot{\xi}^2+\omega_1^2\xi^2\Big) +
    \frac{\beta}{2}\Big(\dot{\eta}^2+\omega_2^2\eta^2\Big)\,,
    \end{equation}
with $\alpha,\beta$ being arbitrary real constants. Using
(\ref{xieta}), we can write $E_{\alpha,\beta}$ as a quadratic form
of $\phi$ and its derivatives up to the third order:
\begin{equation}\label{EPphi}\begin{array}{ll}\displaystyle
E_{\alpha,\beta}&
\displaystyle=\frac{\alpha}{2}\left[\Big(\frac{\dddot{\phi}+\omega_2^2\dot{\phi}}{\omega_2^2-\omega_1^2}\Big)^2+
\omega_1^2\Big(\frac{\ddot{\phi}+\omega_2^2\phi}{\omega_2^2-\omega_1^2}\Big)^2\right]+
\frac{\beta}{2}\left[\Big(\frac{\dddot{\phi}+\omega_1^2\dot{\phi}}{\omega_1^2-\omega_2^2}\Big)^2+
\omega_2^2\Big(\frac{\ddot{\phi}+\omega_1^2\phi}{\omega_1^2-\omega_2^2}\Big)^2\right]
\\[7mm]\displaystyle &
\displaystyle=\frac{\alpha A_1^2\omega_1^2}{2}+\frac{\beta
A_2^2\omega_2^2}{2}.
\end{array}\end{equation}
 If $\alpha\beta\neq 0$, then the only critical point of the function $E_{\alpha,\beta}(\phi,\dot\phi,\ddot\phi,\dddot\phi)$ is zero. The quadratic form $E_{\alpha,\beta}$ is positive definite whenever $\alpha>0$ and $\beta>0$.
 The last fact ensures the boundedness of motion for any choice of initial
 data\footnote{In this simple case, the explicit solution (\ref{phi=xi+eta}), (\ref{osc}) makes obvious that the motion is
 bounded. In many cases, the positive definite integral can be known, while the explicit solutions are unknown.}.

In general, we say that the classical dynamics is \textit{stable} in
a vicinity of a phase-space point $\phi_0$, if $\phi_0$ provides a
local minimum for a conserved quantity $E$ and the Hessian matrix
$d^2E$ is positive definite at $\phi_0$. In this case the level
surfaces $E=E_0$, where $E_0$ is close enough to the minimum value,
are compact and the motion is bounded in the phase space. In the
subsequent discussion we will call a conserved quantity $E$ positive
definite (in a vicinity of its extremum point $\phi_0$) if so is its
Hessian matrix $d^2E$.

In the case of PU oscillator we have the two-parameter family
(\ref{EPphi}) of conserved quantities and at least two physically
reasonable candidates for the energy. First of all, as we are
dealing with the pair of oscillations (\ref{osc}), it is quite
natural to define the energy of the PU model as the total energy of
two uncoupled harmonic oscillators, namely, $$E_{1,1} =
\frac{A_1^2\omega_1^2}{2}+\frac{A_2^2\omega_2^2}{2}.$$ This energy
is positive definite and its conservation ensures the classical
stability of the PU oscillator.

Another possibility is suggested by the Noether theorem \cite{K-S}.
In Lagrangian mechanics the canonical energy is defined as the
integral of motion corresponding to the invariance of a conservative
system under the time translations. This correspondence, being
applied to the PU oscillator, leads to an unbounded energy as we
explain below.

The time derivative of any integral of motion $E$ is to be
proportional to the l.h.s. of equations of motion, i.e.,
\begin{equation}\label{dEdt}
\frac{dE}{dt}=Q \frac{\delta S}{\delta \phi}\,.
\end{equation}
The coefficient $Q=Q(\phi,\dot{\phi},\ddot{\phi}, \dddot{\phi})$ is
called the \textit{characteristic} of the conserved quantity $E$.
The Noether theorem connects the integrals of motion to the
symmetries of action by identifying the characteristic $Q$ with the
infinitesimal symmetry transformation:
\begin{equation}\label{Qtransf}
    \delta_\varepsilon \phi = \varepsilon Q \, , \quad \delta_\varepsilon S=0\qquad
    \Leftrightarrow\qquad
    Q\frac{\delta S}{\delta \phi}=\frac{dE}{dt}
\end{equation}
for some $E=E(\phi,\dot\phi,\ddot\phi,\dddot\phi)$. In this way, the
invariance of the action (\ref{PUL}) with respect to the time
translation $\delta_\varepsilon \phi= -\dot{\phi}\varepsilon$ gives
rise to the Noether energy (\ref{NE}). On the other hand, one can
find the following expression for the characteristic of the
conserved quantity (\ref{EPphi}):
\begin{equation}\label{CPU}
Q_{\alpha,\beta}=\frac{(\alpha+\beta)\dddot{\phi}+(\alpha\omega_2^2+\beta\omega_1^2)\dot{\phi}}
{\omega_1^2-\omega_2^2}\,\,.
\end{equation}
So, the identification $Q=-\dot\phi$ implies that $\alpha=-\beta=1$
and the corresponding Noether energy reads
\begin{equation}\label{NEPU}
E_{1,-1}=\frac{2\dddot{\phi}\dot{\phi}-(\ddot{\phi})^2+(\omega_1^2+\omega_2^2)\dot{\phi}^2+
\omega_1^2\omega_2^2\phi^2}{2(\omega_2^2-\omega_1^2)}=\frac{A_1^2\omega_1^2}{2}-\frac{A_2^2\omega_2^2}{2}.
\end{equation}
Unlike $E_{1,1}$, this energy is not positive definite. The positive
definite integrals of motion
 (\ref{EPphi}) correspond to $\alpha>0$, $\beta>0$ and their characteristics
 (\ref{CPU}) are bound to involve the third derivative of $\phi$.  As a result,
the usual Noether theorem can't connect a positive conserved
quantity to the time translation.

A more general correspondence between symmetries and integrals of
motion is established by means of the Lagrange anchor
\cite{KapLS2010}, see also Appendix B. The Lagrange anchor is a
differential operator that satisfies certain compatibility
conditions with the equations of motion, see the definition
(\ref{anchor}). Given equations of motion, the Lagrange anchor is
not necessarily unique and the different Lagrange anchors establish
different connections between symmetries and conservation laws. In
particular, for the PU oscillator we have the two-parameter family
of the Lagrange anchors (\ref{App2VPU}):
\begin{equation}\label{VPU1}
V_{\rho,\sigma}=
\frac{\rho}{\omega_2^2-\omega_1^2}\left(\frac{d^2\phantom{t}}{dt^2}+\omega_2^2\right)+
    \frac{\sigma}{\omega_1^2-\omega_2^2}\left(\frac{d^2\phantom{t}}{dt^2}+\omega_1^2\right)\,,
\end{equation}
with $\rho$ and $\sigma$ being arbitrary real constants. The details
about deriving this Lagrange anchor are collected in Appendix
\ref{AppC}.

Each Lagrange anchor maps characteristics to symmetries by the rule
(\ref{NT}). Applying the Lagrange anchor (\ref{VPU1}) to the
characteristic (\ref{CPU}), we get the following symmetry that
corresponds to the integral of motion (\ref{EPphi}):
\begin{equation}\label{CSPU}\begin{array}{ll}
    \displaystyle\delta_\varepsilon\phi=\varepsilon V_{\rho,\sigma}(Q_{\alpha,\beta})&=\displaystyle \frac{\varepsilon}{(\omega_1^2-\omega_2^2)^2}
    \Big[(\alpha+\beta)(\rho-\sigma)\phi^{(5)}+
    (\omega_1^2(\alpha\rho+2\beta\rho-\beta\sigma)\\[3mm] & \left. \displaystyle-
    \omega_2^2(\beta\sigma+2\alpha\sigma-\alpha\rho))\dddot{\phi}+
    (\beta\rho\omega_1^4+(\alpha\rho-\beta\sigma)\omega_1^2\omega_2^2-
    \alpha\sigma\omega_2^4)\dot\phi\right]\,.
\end{array}\end{equation}

Let us consider this relationship from the perspective of having
alternative integrals of motion connected with the time translation.
To establish the correspondence, we re-arrange (\ref{CSPU}) to
absorb the higher derivative term with $\phi^{(5)}$ by the equation
of motion \footnote{The symmetry transformation is defined modulo
on-shell vanishing terms. Once the equation is of fourth order, the
fourth and higher derivatives can always be excluded from the
symmetry transformation. In particular, the fifth derivative in
(\ref{CSPU}) may be included into on-shell vanishing terms.}:
\begin{equation}\label{CSPU5}
\begin{array}{rll}
   \displaystyle  \delta_{\varepsilon}\phi&\displaystyle=&\displaystyle\varepsilon\frac{(\omega_1^2-\omega_2^2)(\alpha\sigma+\beta\rho)\dddot{\phi}+
    (\beta\rho\omega_1^4+(\alpha\sigma-\beta\rho)\omega_1^2\omega_2^2-
    \alpha\sigma\omega_2^2)\dot\phi}{(\omega_1^2-\omega_2^2)^2}\\[5mm]\displaystyle&\displaystyle+&\displaystyle
    \varepsilon\frac{(\alpha+\beta)(\rho-\sigma)}{\omega_1^2-\omega_2^2}\frac{d}{dt}\frac{\delta
    S}{\delta\phi}\,.
    \end{array}
\end{equation}
The anchor connects the general characteristic (\ref{CPU}) with the
time translation $\delta_\varepsilon\phi=-\dot{\phi}\varepsilon$ if
the coefficient at $\dddot\phi$ vanishes. This leads to the
condition $\alpha\rho+\beta\sigma=0$. The correct coefficient at the
first derivative is provided by $\alpha\rho=1$. Solving these
conditions for $\rho$ and $\sigma$, we see that the Lagrange anchor
$V_{\frac1\alpha,-\frac1\beta}$
connects the general nondegenerate integral of motion (\ref{EPphi})
to the time translation
\begin{equation}\label{deltax}
\delta_\varepsilon\phi=\varepsilon
V_{\frac1\alpha,-\frac1\beta}(Q_{\alpha,\beta})=-\varepsilon
\dot{\phi}-\frac{(\alpha+\beta)^2}{\alpha\beta}
\frac{\varepsilon}{\omega_1^2-\omega_2^2}\frac{d}{dt}\frac{\delta
S}{\delta \phi}\,.
\end{equation}
We have observed above that any integral of motion (\ref{EPphi})
with $\alpha\beta\neq 0$ can be connected to the time translation by
specification of the free parameters in the general Lagrange anchor
(\ref{VPU1}). The Noether energy (\ref{NEPU}) is mapped to the
symmetry by the canonical Lagrange anchor. The positive integrals of
motion are mapped to the generator of time translations by the
non-canonical Lagrange anchors (\ref{VPU1}) with
$\rho>0\,,\sigma<0$.

Let us stress once and again that different Lagrange anchors result
in different quantizations of one and the same classical system (see
Appendix \ref{AppA} and \cite{KazLS, LS1}). For the first-order
ODEs, a Lagrange anchor always defines\footnote{under the additional
assumption of integrability, see \cite{KapLS2010}. The field
independent Lagrange anchors are automatically integrable.} a
Poisson bracket on the phase space of the system, while the
corresponding energy becomes a Hamiltonian \cite{KazLS,KapLS2013}.
Once the equations of motion admit several Lagrange anchors, they
admit several Poisson brackets and Hamiltonians. If the Hamiltonian
is positive, one can expect the bounded spectrum of energy and
quantum stability, while the unbounded energy usually results in
quantum instability. So, the choice of the Lagrange anchor and the
energy gains importance when the quantum stability is concerned.

We do not elaborate here on the generalities of the connection
(which is basically one-to-one for ODE's, modulo certain equivalence
relations) between the integrable Lagrange anchors and the Poisson
brackets, see \cite{KazLS, BG2011a, KapLS2013}. We will just
explicitly demonstrate that any non-degenerate integral of motion
(\ref{EPphi}) leads to the corresponding Hamiltonian form of
dynamics.

Consider the Hamiltonian formulation for the model (\ref{PUL}).
Following the Ostrogradsky method, we introduce the canonical
variables
\begin{equation}\label{cvPU}
    q_1=\phi, \quad q_2=\dot{\phi}, \quad p_1= \frac{\partial L}{\partial \dot{\phi}}-\frac{d}{dt}\frac{\partial L}{\partial
    \ddot{\phi}}=-\frac{2\dddot{\phi}+(\omega_1^2+\omega_2^2)\dot{\phi}}{2(\omega_1^2-\omega_2^2)} , \quad p_2=\frac{\partial
    L}{\partial\ddot{\phi}} =\frac{2\ddot{\phi}+(\omega_1^2+\omega_2^2)\phi}{2(\omega_1^2-\omega_2^2)}\, ,
\end{equation}
which have the canonical Poisson brackets
\begin{equation}\label{PUPB}
    \{q_i,p_j\}_O=\delta_{ij}\, , \qquad \{q_i,q_j\}_O=\{p_i,p_j\}_O=0,
    \quad i,j=1,2.
\end{equation}
Then $\phi,\dot{\phi}, \ddot{\phi}, \dddot{\phi}$ can be expressed
in terms of the phase space variables:
\begin{equation}\label{dPU}
    \phi=q_1, \quad \dot{\phi}=q_2, \quad \ddot{\phi} =(\omega_1^2-\omega^2_2)p_2-\frac12(\omega_1^2+\omega_2^2)q_1\,,
    \quad\dddot{\phi}=(\omega_2^2-\omega_1^2)p_1-\frac12(\omega_1^2+\omega_2^2)q_2
    \ .
\end{equation}
The Ostrogradsky Hamiltonian, being the phase-space expression for
Noether's energy (\ref{NEPU}),  reads
\begin{equation}\label{OH}
H_O=p_1q_2-\frac{\omega_1^2+\omega_2^2}{2}p_2q_1+
\frac{\omega_1^2-\omega_2^2}{2}\Big(p_2^2+\frac14{q_1^2}\Big)\,.
\end{equation}
The phase space variables $z^I=\{q_1,q_2,p_1,p_2\}$ satisfy the
Hamiltonian equations
\begin{equation}\label{HE}
\dot{z}^I=\{z^I,H_O\}_O\,
\end{equation}

Because of the aforementioned correspondence between the Lagrange
anchors in mechanical systems and Poisson structures, the
two-parameter set of Lagrange anchors (\ref{VPU1}) and the energy
functions (\ref{EPphi}) imply the existence of two-parameter sets of
Poisson brackets and Hamitonians. These read
\begin{equation}\label{VPB}
    \begin{array}{lll}
    \displaystyle\{q_1,q_2\}_{\alpha,\beta}=\frac{1}{\alpha}+\frac{1}{\beta}\,,\qquad&
    \displaystyle\{q_1,p_1\}_{\alpha,\beta}=\frac{1}{2}\Big(\frac{1}{\alpha}-\frac{1}{\beta}\Big)\,, \qquad &
    \displaystyle\{q_1,p_2\}_{\alpha,\beta}=0\,,\\[3mm]
    \displaystyle\{q_2,p_1\}_{\alpha,\beta}=0\,,\qquad&
    \displaystyle\{q_2,p_2\}_{\alpha,\beta}=\frac{1}{2}\Big(\frac{1}{\alpha}-\frac{1}{\beta}\Big)\,,\qquad&
    \displaystyle\{p_1,p_2\}_{\alpha,\beta}=\frac{1}{4}\Big(\frac{1}{\alpha}+\frac{1}{\beta}\Big)\,,
\end{array}
\end{equation}
\begin{equation}\label{HPU}
H_{\alpha,\beta}=\frac{\alpha}{2}\Big[(p_1+q_2/2)^2+\omega_1^2(p_2-q_1/2)^2\Big]+
\frac{\beta}{2}\Big[(p_1-q_2/2)^2+\omega_2^2(p_2+q_1/2)^2\Big]\,.
\end{equation}
The Hamiltonians $H_{\alpha , \beta}$ are derived from $E_{\alpha ,
\beta}$ by substitution $\phi, \dot{\phi}, \ddot{\phi},
\dddot{\phi}$ in terms of the phase-space variables (\ref{dPU}). The
Ostrogradsky Hamiltonian and bracket correspond to $\alpha=1,
\beta=-1$:
$$
\{\;\cdot\;,\;\cdot\;\}_O=\{\;\cdot\;,\;\cdot\;\}_{1,-1}\,,\qquad
H_O=H_{1,-1}\,.
$$
Notice that the brackets and Hamiltonians with different $\alpha ,
\beta$ are not obtained from each other by canonical
transformations. It is an obvious fact because the brackets between
the same variables essentially depend on the parameters. For
example, the original coordinate $q_1=\phi$ Poisson commutes with
the velocity $q_2=\dot{\phi}$ once $\alpha=-\beta$, while they are
conjugate when $\alpha=\beta$; $q_1=\phi$ is a conjugate to
$p_1=-\frac{2\dddot{\phi}+(\omega_1^2+\omega_2^2)\dot{\phi}}{2(\omega_1^2-\omega_2^2)}$
with respect to the bracket (\ref{VPB}) once $\alpha=-\beta$, while
they commute when $\alpha=\beta$. However, for any $\alpha, \beta$,
the corresponding Hamiltonian equations with the brackets $\{ \cdot
, \cdot \}_{\alpha , \beta}$ and the Hamiltonians $H_{\alpha ,
\beta}$ coincide with each other, and in particular with the
Ostrogradsky system, i.e.,
\begin{equation}\label{HEV}
\dot{z}^I=\{z^I,H_{\alpha,\beta}\}_{\alpha,\beta}\equiv\{z^I,
H_O\}_O\, , \qquad \forall\alpha\neq 0,\qquad\forall\beta\neq 0\,.
\end{equation}
Thus, the phase space equations of the PU oscillator admit a
two-parameter set of brackets and Hamiltonians.

For $\alpha> 0$, $\beta>0$ (that corresponds to $H_{\alpha ,
\beta}>0$) the special coordinates can be introduced
\begin{equation}\label{CanVar}\begin{array}{ll}
    \displaystyle\pi_\xi=\sqrt{\alpha}(p_1+q_2/2)\,,&\qquad\displaystyle
    \chi_\xi\equiv\sqrt{\alpha}\xi=\sqrt{\alpha}(q_1/2-p_2)\,,\\[3mm]
    \displaystyle\pi_\eta=\sqrt{\beta}(q_2/2-p_1)\,,&\qquad\displaystyle
    \chi_\eta\equiv\sqrt{\beta}\eta=\sqrt{\beta}(p_2+q_1/2)\,.
\end{array}\end{equation}
In these coordinates, the brackets (\ref{VPB}) take the canonical
form
\begin{equation}\label{PUPBV}
    \{\chi_i,\pi_j\}_{\alpha,\beta}=\delta_{ij}\, , \qquad \{\chi_i,\chi_j\}_{\alpha,\beta}=
    \{\pi_i,\pi_j\}_{\alpha,\beta}=0,\quad i,j=\xi,\eta.
\end{equation}
The Hamiltonian (\ref{HPU}) reduces to that of a two-dimensional
harmonic oscillator, namely,
\begin{equation}\label{HPUCanVar}
H_{\alpha,\beta}=\frac{1}{2}\big({\pi_\xi^2}+\omega_1^2\chi_\xi^2\big)+
\frac{1}{2}\big({\pi_\eta^2}+\omega_2^2\chi_\eta^2\big)\,.
\end{equation}
If the PU oscillator is quantized with the Hamiltonian  (\ref{HPU})
 by imposing the commutation relations according to the corresponding bracket (\ref{VPB}) with $\alpha > 0,
\ \beta>0$, this is equivalent to canonical quantization with the
canonical bracket (\ref{PUPBV}) and Hamiltonian (\ref{HPUCanVar}).
This means that the quantum theory with the non-canonical Lagrange
anchor leads to the positive energy spectrum, while the canonical
choice results in the spectrum unbounded from below .

Let us summarize the conclusions made in this section that apply (as
we will see in the next sections) to a wide class of
higher-derivative dynamics. Once the free higher-derivative system
admits factorization, it turns out classically stable, because the
two-parameter family exists of the conserved quantities that
includes the bounded functions. The model was shown to admit a
two-parameter family of the Lagrange anchors that connect the
conserved quantities with the symmetry of system under time
translation. This allows one to consider any of the integrals as the
energy. As we have seen, the diversity of the Lagrange anchors
admitted by the higher-derivative dynamics makes possible to choose
between inequivalent quantizations. It turns out that the classical
stability can be retained at the quantum level by appropriate choice
of the Lagrange anchor.

In the next section, we generalize these observations to a broad
class of interacting higher-derivative systems. The example of the
interaction that does not break the stability of the PU oscillator
will be provided. Then, in Section 4, we will consider examples of
stability in higher-derivative field theories.

\section{Nonlinear factorization}
In this section, we formulate the general pattern for factorizing
not necessarily linear higher-derivative systems. This pattern can
be seen in its simplest form already from the example of the PU
oscillator. Once the higher-derivative dynamics is factorized in
this sense, the stability turns out to be a common occurrence as
much as it happens in the usual dynamics without higher derivatives.
As we will demonstrate, many of higher-derivative systems of this
class appear to be stable, though their canonical energy is
unbounded from below.

Suppose that $\xi$, $\eta$, and  $\phi$ are $n$-component fields on
space-time with local coordinates $\{x^\mu\}$. Given $n\times n$
matrix differential operator $\mathcal{P}$, define $\mathcal{Q}$ by
the relation\footnote{This relation can be relaxed in various ways.
For example $P+Q$ is sufficient to be an invertible matrix
differential operator, not necessarily unit, if $\mathcal P$ and
$\mathcal Q$ commute.}
\begin{equation}\label{PU}
  1=\mathcal{P}+\mathcal{Q}\,.
\end{equation}
 Clearly,  $[\mathcal{P}, \mathcal{Q}]=0$.
 Using these operators and an arbitrary vector-valued non-linear differential operator $\mathcal{F}$,
 we can define two systems of field equations. The first one includes two groups of equations
 \begin{equation}\label{ODE1}
  \mathcal{P}\xi+ \mathcal{F}(\xi, \eta)=0\,,\qquad \mathcal{Q}\eta + \mathcal{F}(\xi,\eta)=0\,,
 \end{equation}
 while the second is given by
 \begin{equation}\label{ODE2}
 \mathcal{P}\mathcal{Q}\phi + \mathcal{F}(\mathcal{Q}\phi,\mathcal{P}\phi)=0\,.
 \end{equation}
It is easy to check that the relations
\begin{equation}\label{map}
  \xi=\mathcal{Q}\phi\,,\qquad \eta=\mathcal{P}\phi\,,\qquad  \phi=\xi+\eta\,
\end{equation}
establish a one-to-one correspondence between solutions of both the
systems. So, the systems (\ref{ODE1}) and (\ref{ODE2}) are
equivalent and may be thought of as two different representations of
one and the same theory. We will refer to  them as $\xi\eta$- and
$\phi$-representations. The PU oscillator provides the simplest
example of factorization with $\mathcal{F}=0$, cf. (\ref{LEPU}),
(\ref{phi=xi+eta}), (\ref{PUxieta}).

The $\xi\eta$-representation (\ref{ODE1}) may be viewed as a special
way to depress the order of system (\ref{ODE2}). For example, if
$\mathcal P$ is of the second order, and $\mathcal{F}$ is algebraic,
then the fourth-order equations (\ref{ODE2}) are equivalent to the
second-order equations (\ref{ODE1}). The operator $\mathcal{F}$ can
be considered as an interaction included\footnote{The consistency of
the interaction is not granted by this construction. We suppose the
interaction is consistent, and study stability. For detailed
discussion of consistency of interaction in the non-Lagrangian
context we refer the reader to \cite{KapLS2013i}.} into the free
system $\mathcal{P}\mathcal{Q}\phi=0$. In this way, the
factorization can still be efficient for keeping track of stability
in the interacting higher-derivative dynamics.

Let us assume that $\mathcal{P}^\dag=\mathcal{P}$ and construct
$\mathcal{F}(\xi,\eta)$ in the following way. Given a function
$U(\phi,\partial\phi,
\partial^2\phi, \ldots,\partial^N\phi )$, consider it's Euler-Lagrange derivative for
brevity denoted by
$$U'=\sum_{k=0}^N(-1)^k\frac{\partial^k}{\partial
x^{\mu_{1}}\ldots \partial x^{\mu_{k}}} \frac{\partial
U}{\partial(\partial_{\mu_{1}}\ldots\partial_{\mu_{k}}\phi)}\,.
$$
The nonlinearity $\mathcal{F}$ in (\ref{ODE1}) can be chosen as
\begin{equation}\label{FU'}
\mathcal{F}(\xi,\eta)=-U'|_{\phi\rightarrow \alpha\xi -\beta\eta},
\end{equation}
 with
$\alpha$ and $\beta$ being nonzero constants. Then system
(\ref{ODE1}) comes from the least action principle for
\begin{equation}\label{L}
  S_1 \ [\xi(x),\eta(x)] =\int L_1 dx,\qquad
  L_1=\frac{\alpha}{2}\xi\mathcal{P}\xi-\frac{\beta}{2}\eta\mathcal{Q}\eta-U(\alpha\xi-\beta\eta)\,
  ,
\end{equation}
while equations (\ref{ODE2}) are not necessarily variational. For
the special nonlinearity (\ref{FU'}), the equations (\ref{ODE1})
take the form
\begin{equation}\label{ODE1L}
\frac{\delta
S_1}{\delta\xi}\equiv\alpha(\mathcal{P}\xi-U'(\alpha\xi-
\beta\eta))=0\,,\qquad\frac{\delta
S_1}{\delta\eta}\equiv-\beta(\mathcal{Q}\eta-U'(\alpha\xi-
\beta\eta))=0\, ,
\end{equation}
 and (\ref{ODE2}) read
\begin{equation}\label{ODE2L}
\mathcal{P}\mathcal{Q}\phi-
U'(\alpha\mathcal{Q}\phi-\beta\mathcal{P}\phi)=0\, .
\end{equation}

In some cases, the dynamical equations (\ref{ODE1L}) and
(\ref{ODE2L}) should be multiplied by an overall dimensional
constant to ensure the proper dimension of the action (\ref{L}). For
example, for the PU oscillator (\ref{PUL}), it is convenient to take
this factor as $\omega_2^2-\omega_1^2$. Once the dimensional
coefficient is introduced, all the expressions in this section for
the actions, equations of motion and conserved currents are to be
multiplied by this constant, while the characteristics, symmetries
and Lagrange anchors remain intact. As the dimensional coefficient
adds no essential generality, but complicates the explicit
expressions, it is omitted from most of expressions.

The least action principle for (\ref{ODE1L}) not necessarily makes
equations (\ref{ODE2L}) Lagrangian. The obvious variational vertex
$\mathcal{F}(\mathcal{P}\phi,\mathcal{Q}\phi)=-U'(\phi)$ corresponds
to the special choice of constants $\alpha=-\beta=1$. The
corresponding action reads
\begin{equation}\label{L2}
S_2[\phi(x)]=\int L_2 dx\,, \qquad L_2=\frac{1}{2}\phi
\mathcal{P}\mathcal{Q}\phi-U(\phi)\,.
\end{equation}

If the action (\ref{L}) is invariant under the space-time
translations $x^\mu \rightarrow x^\mu-\varepsilon^\mu$ then (by the
Noether theorem (\ref{Qtransf})) the system of equations
(\ref{ODE1L}) admits the conserved current $J(\xi,\eta)$ such that
\begin{equation}\label{Char2}
    \partial_\mu J^\mu=-\varepsilon^\mu\partial_\mu\xi\frac{\delta
    S_1}{\delta\xi}-\varepsilon^\mu\partial_\mu\eta\frac{\delta
    S_1}{\delta\eta}\,.
\end{equation}
It is expressible through the canonical energy-momentum tensor as
\begin{equation}\label{EMC}
    J^\mu=\Theta^\mu_{\phantom{a}\nu}\varepsilon^\nu\,,
\end{equation}
where
\begin{equation}\label{EMT}
    \Theta^\mu_{\phantom{a}\nu}(\xi,\eta)=\sum_{\phi=\xi,\eta}\sum_{k=1}^{N}\Big[(\partial_{\mu_1}\ldots\partial_{\mu_{k-1}}\partial_\nu\phi)
    \sum_{m=k}^{N}(-1)^{(m-k)}\partial_{\mu_k}\ldots\partial_{\mu_{m-1}}\frac{\partial
    L_1}{\partial
    (\partial_{\mu_1}\ldots\partial_{\mu_{m-1}}\partial_\mu\phi)}\Big]-\delta^\mu_\nu
    L_1\,.
\end{equation}
Here, the sums by $k$ and $m$ run up to the maximal order of
derivatives $N$ entering the Lagrangian (\ref{L}). The
energy-momentum tensor is given by the sum
\begin{equation}\label{EMTL1}
    \Theta^\mu_{\phantom{a}\nu}(\xi,\eta)=\alpha(\Theta_\mathcal{P}){}^\mu_{\phantom{a}\nu}(\xi)-
    \beta(\Theta_\mathcal{Q}){}^\mu_{\phantom{a}\nu}(\eta)+(\Theta_U){}^\mu_{\phantom{a}\nu}(\xi,\eta)\,,
\end{equation}
where $(\Theta_\mathcal{P}){}^\mu_{\phantom{a}\nu}$ and
$(\Theta_\mathcal{Q}){}^\mu_{\phantom{a}\nu}$ are the
energy-momentum tensors for the Lagrangian free theories
$\mathcal{P}\xi=0$ and $\mathcal{Q}\eta=0$, while the term
$(\Theta_U){}^\mu_{\phantom{a}\nu}$ is the energy-momentum tensor of
``interaction''. By construction, the component
$\Theta^0_{\phantom{a}0}$ has the sense of the energy density of the
theory (\ref{ODE1L}), so that the total energy of the system is
given by the integral $E=\int_{\text{space}}
\Theta^0_{\phantom{a}0}$. The stability of the theory (\ref{ODE1L})
is provided by the condition $\Theta^0_{\phantom{a}0}\geq 0$.

An alternative analysis of stability can be done by switching to the
Hamiltonian formalism for the theory (\ref{L}). The stability of the
theory (\ref{ODE1L}) is guaranteed if the Hamiltonian $H=E$ is
positive definite. This approach may be convenient for the theories
whose lower-order Lagrangian formulations (\ref{L}) are
well-studied. As an example we can mention the conformal higher-spin
fields \cite{METSAEV2014}.

Let us now prove that  in the $\phi$-representation the
energy-momentum tensor (\ref{EMT}) is also associated with the
space-time translations. This tensor can ensure stability of the
theory (\ref{ODE2L}) much like the canonical energy-momentum tensor
does in the usual theory without higher derivatives. Substituting
$\phi$ into (\ref{EMC}) by the rule (\ref{map}), we find that the
tensor
$\Theta^\mu_{\phantom{a}\nu}(\mathcal{Q}\phi,\mathcal{P}\phi)$
conserves,
\begin{equation}\label{EMC2}
\partial_\mu
\Theta^\mu_{\phantom{a}\nu}(\mathcal{Q}\phi,\mathcal{P}\phi)=
\big[\partial_{\nu}(\beta\mathcal{P}-\alpha\mathcal{Q})\phi\big]
\,\big[\mathcal{P}\mathcal{Q}\phi-U'(\alpha\mathcal{Q}\phi-\beta\mathcal{P}\phi)\big]\,,
\end{equation}
and the corresponding characteristic reads
\begin{equation}\label{EMC2char}
    Q_\nu=\partial_\nu(\beta\mathcal{P}-\alpha\mathcal{Q})\phi\,.
\end{equation}
Obviously, $\Theta^0_{\phantom{a}0}(\xi,\eta)\geq0$ implies
$\Theta^0_{\phantom{a}0}(\mathcal{Q}\phi,\mathcal{P}\phi)\geq0$

Notice that the order of variational equations (\ref{ODE1L}) may be
lower than the order of equations (\ref{ODE2L}). By this reason, the
use of variational formulation (\ref{L}) allows one to surpass the
obstructions to the existence of positive definite energy in
theories with higher derivatives. For example, if the differential
operators $\mathcal{P}$ and $\mathcal{Q}$ are of the second order,
then the positive definite energy density may exist even if the
theory (\ref{ODE2L}) is nonsingular. On the other hand, the use of
the Noether theorem for the constriction of conservation laws sets
the natural upper bound for the order of action (\ref{L}). This
suggests to concentrate on the theories (\ref{ODE2L}) for which the
operators $\mathcal{P}$, $\mathcal{Q}$ are at most of the second
order and $U=U(\phi,\partial\phi)$ depends on at most first
derivatives of the field. However, if the higher-derivative models
(\ref{L}) with the positive definite Noether energy are found in the
future, our construction will be applicable to them as well.

More information about stability of the theory (\ref{ODE2L}) may be
obtained  if the structure of the energy-momentum tensor (\ref{EMT})
is taken into account. For example, if the two factors are stable
(i.e., $\alpha(\Theta_\mathcal{P}){}^0_{\phantom{a}0},
-\beta(\Theta_\mathcal{Q}){}^0_{\phantom{a}0}\geq0$ for some values
of $\alpha$ and $\beta$) and $(\Theta_U){}^0_{\phantom{a}0}\geq0$,
the theory (\ref{ODE1}) is stable. This fact can be used for a
systematical constriction of stable interacting higher-derivative
theories. If both the factors are stable, but the interaction term
is not positive definite, the energy can still have a local minimum
in a neighborhood of zero solution. Such theories with ``locally
stable'' behavior are also considered as physically acceptable
models. They can be studied within the perturbation theory. The
examples are known of the locally stable models with not necessarily
positive energy \cite{Smilga2005,SMILGA2014,PAVSIC2013,PAVSIC2013b}.
In such theories with ``benign ghosts'' we can expect the existence
of (yet unknown) Lagrange anchor and an alternative positive
definite conserved energy. In other cases, the stability of a theory
cannot be guaranteed even in a small neighborhood of the vacuum
solution. The theories of this type are branded as having
``malicious ghosts'' \cite{Smilga2005} and cannot be considered as
physical.

Whenever the system of equations (\ref{ODE2L}) is not variational,
the relationship between the conserved tensor (\ref{EMTL1}) and the
space-time translations can be established by the Lagrange anchor.
In Appendix D we find that for factorable systems the Lagrange
anchor reads:
\begin{equation}\label{V2}
  V=\frac{1}{\alpha}\mathcal{Q}-\frac{1}{\beta}\mathcal{P}+
  \frac{(\alpha+\beta)^2}{\alpha\beta}U''(\alpha\mathcal{Q}\phi-\beta\mathcal{P}\phi)\,.
\end{equation}
The action of the matrix differential operator $U''$ on an arbitrary
characteristic $Q(\phi(x))$ is defined by
\begin{equation}\label{}
  U''(\phi)Q=\int dx\frac{\delta
  U'(\phi)}{\delta\phi(x)}Q(\phi(x))\,.
\end{equation}
Verification of the defining property (\ref{anchor}) for the
Lagrange anchor (\ref{V2}) requires some technical details provided
in Appendix \ref{AppD}. Applying (\ref{V2}) to the characteristic
(\ref{EMC2char}), we get the space-time translation symmetry

\begin{equation}\begin{array}{l}
\displaystyle \delta_\varepsilon\phi=\varepsilon^\nu V
(Q_\nu)=\varepsilon^\nu\left(\frac{1}{\alpha}\mathcal{Q}-\frac{1}{\beta}\mathcal{P}+
\frac{(\alpha+\beta)^2}{\alpha\beta}U''(\alpha\mathcal{Q}\phi-\beta\mathcal{P}\phi)\right)
(\beta\mathcal{P}-\alpha\mathcal{Q})\partial_\nu\phi=
\\[3mm]\displaystyle=
\left(\frac{1}{\alpha}\mathcal{Q}-\frac{1}{\beta}\mathcal{P}\right)
(\beta\mathcal{P}-\alpha\mathcal{Q})\varepsilon^\nu\partial_\nu\phi-
\frac{(\alpha+\beta)^2}{\alpha\beta}U''(\alpha\mathcal{Q}\phi-\beta\mathcal{P}\phi)
\varepsilon^\nu\partial_\nu(\alpha\mathcal{Q}\phi-\beta\mathcal{P}\phi)=\\[3mm]
\displaystyle=-\varepsilon^\nu\partial_\nu
\phi+\frac{(\alpha+\beta)^2}{\alpha\beta}\varepsilon^\nu\partial_\nu\big(\mathcal{Q}\mathcal{P}
\phi-U'(\alpha\mathcal{Q}\phi-\beta\mathcal{P}\phi)\big)\approx
-\varepsilon^\nu\partial_\nu \phi\,.
\end{array}\end{equation}
This relation allows us to identify  the conserved current (\ref{EMC2})
with the energy-momentum current of the theory (\ref{ODE2L}).

Let us illustrate the general construction above by the example of
PU oscillator. The operators $\mathcal{P}$ and $\mathcal{Q}$ now
take the form
\begin{equation}\label{PQPU}
\mathcal{P}=\frac{1}{\omega_1^2-\omega_2^2}\left(\frac{d^2}{dt^2}+\omega^2_1\right),\qquad
\mathcal{Q}=\frac{1}{\omega_2^2-\omega_1^2}\left(\frac{d^2}{dt^2}+\omega^2_2\right)\,
\, . \end{equation} Upon substituting (\ref{PQPU}) into
(\ref{ODE2L}) and multiplying by the overall factor
$\omega^2_2-\omega^2_1$, we get the following  equation of motion:
\begin{equation}\label{PUint}
T\equiv\frac{1}{\omega_1^2-\omega_2^2}\left(\frac{d^2}{dt^2}
+\omega_1^2\right)\left(\frac{d^2}{dt^2}+\omega_2^2\right)\phi-
U'\Big(\frac{(\alpha+\beta)\ddot{\phi}+(\alpha\omega_2^2+\beta\omega_1^2)\phi}
{\omega_2^2-\omega_1^2}\Big)=0\,.
\end{equation}
For simplicity sake we assume the function  $U(\phi)$ to depend on $\phi$ but not on its derivatives, so that $U'=d U(\phi)/d\phi$. The two-parameter family
of integrals of motion reads
\begin{equation}\label{PUEint}
E=E_{\alpha,\beta}+U\Big(\frac{(\alpha+\beta)\ddot{\phi}+(\alpha\omega_2^2+\beta\omega_1^2)\phi}
{\omega_2^2-\omega_1^2}\Big)\,,
\end{equation}
where $E_{\alpha,\beta}$ is defined by (\ref{EPphi}). One can easily check that
\begin{equation}\label{PUChar}
\frac{dE}{dt}=Q T\,,\qquad
Q=\frac{(\alpha+\beta)\dddot{\phi}+(\alpha\omega_2^2+\beta\omega_1^2)\dot{\phi}}
{\omega_1^2-\omega_2^2}\,.
\end{equation}
Expression (\ref{PUEint}) is positive definite whenever
$\alpha,\beta>0$ and $U\geq0$. In that case the motion is bounded
for any initial data. To the best of our knowledge this is the first
example of the self-interacting PU oscillator whose classical
stability can be proved analytically for all initial data. In the
previously known examples of interactions
\cite{Smilga2005,PAVSIC2013} boundedness of motion has been
demonstrated by numerical computations.

To conclude the consideration of the fourth-order formulation
(\ref{PUint}) let us write out the Lagrange anchor
\begin{equation}\label{LAPU}\begin{array}{ll}
    \displaystyle
    V=&\displaystyle\frac{1}{\alpha}\frac{1}{\omega_2^2-\omega_1^2}\left(\frac{d^2\phantom{t}}{dt^2}+\omega_2^2\right)+
    \frac{1}{\beta}\frac{1}{\omega_2^2-\omega_1^2}\left(\frac{d^2\phantom{t}}{dt^2}+\omega_1^2\right)+
    \\[3mm]
    &\displaystyle+\frac{1}{\omega^2_2-\omega_1^2}\frac{(\alpha+\beta)^2}{\alpha\beta}U''\Big(\frac{(\alpha+\beta)\ddot{\phi}+(\alpha\omega_2^2+\beta\omega_1^2)\phi}
{\omega_2^2-\omega_1^2}\Big)\,,\qquad U''=\frac{d^2
U(\phi)}{d\phi^2}\,,
\end{array}\end{equation}
and the corresponding time-translation symmetry
\begin{equation}\label{deltax1}
\delta_\varepsilon\phi=\varepsilon V(Q)=-\varepsilon
\dot{\phi}-\frac{(\alpha+\beta)^2}{\alpha\beta}
\frac{\varepsilon}{\omega_1^2-\omega_2^2}\frac{dT}{dt}\,.
\end{equation}

The Hamiltonian formulation for the fourth-order theory (\ref{PUint}) can be
derived with the help of the auxiliary action (\ref{L}). In our case, it
takes the form
\begin{equation}\label{S1}
    S_1=\int L_1 dt\,,\qquad L_1=
    \frac{\alpha}{2}(\dot{\xi}^2-\omega_1^2\xi^2)+\frac{\beta}{2}(\dot{\eta}^2-\omega_2^2\eta^2)-
    U(\alpha\xi-\beta\eta)\,.
\end{equation}
Introducing the canonical momenta
\begin{equation}\label{pi}
    p_\xi\equiv\frac{\partial L}{\partial\dot\xi}=\alpha\dot\xi\,,\qquad
    p_\eta\equiv\frac{\partial L}{\partial\dot\eta}=\beta\dot\eta\,,
\end{equation}
and performing the Legendre
transform,  we obtain  the Hamiltonian
\begin{equation}\label{PUH}
    H=\frac{1}{2}\Big(\frac{p_\xi^2}{\alpha}+\alpha\omega_1^2\xi^2\Big)+
    \frac{1}{2}\Big(\frac{p_\eta^2}{\beta}+\beta\omega_2^2\eta^2\Big)+U(\alpha\xi-\beta\eta).
\end{equation}
Obviously, the Hamiltonian (\ref{PUH}) is positive definite simultaneously  with the energy (\ref{PUEint}). The canonical transformation
(\ref{CanVar})
\begin{equation}\label{CTr}
    \pi_\xi=\frac{p_\xi}{\sqrt{\alpha}}\,,\qquad
    \pi_\eta=\frac{p_\eta}{\sqrt{\beta}}\,,\qquad\chi_\xi=\sqrt{\alpha}\xi\,,\qquad
    \chi_\eta=\sqrt{\beta}\eta\,,
\end{equation}
brings the Hamiltonian to the form
\begin{equation}\label{PUH2}
    H=H_{\alpha,\beta}+U(\sqrt\alpha\chi_\xi-\sqrt\beta\chi_\eta)\,.
\end{equation}
As is seen the Hamiltonian (\ref{PUH2}) is a deformation of the free
Hamiltonian (\ref{HPUCanVar}). Quantizing this theory in the usual
way by introducing creation-annihilation operators, we arrive at the
quantum theory with a well-defined ground state and a positive
energy spectrum.

\section{Examples of stable higher derivative field theories}
In this section, we consider two examples of the higher derivative
field theories which  are stable despite the fact that their
canonical energy is unbounded from below. The consideration follows
the general pattern described in the previous section.

\subsection{Scalar field with higher derivatives}
Consider the Lagrangian of a free scalar field $\phi$:
$$
L=\frac{1}{2(m_1^2-m_2^2)}\Big(\Box\phi+m_1^2\phi\Big)\Big(\Box\phi+m_2^2\phi\Big)\,
 ,
$$
where $\Box=\partial_\mu\partial^\mu$ is the D'Alembert operator.
The equation of motion reads
\begin{equation}\label{FSF}
\frac{\delta S}{\delta
\phi}=\frac{1}{m_1^2-m_2^2}\Big(\Box+m_1^2\Big)\Big(\Box+m_2^2\Big)\phi=0\,.
\end{equation}
If $m_1\neq m_2$, the theory has factorable structure (\ref{ODE2})
with the following operators $\mathcal{P}$ and $\mathcal{Q}$:
$$
\mathcal{P}=\frac{\Box+m_1^2}{m_1^2-m_2^2},\qquad\mathcal{Q}=\frac{\Box+m_2^2}{m_2^2-m_1^2}\,.
$$
In the second-order formalism the corresponding fields $\xi$ and
$\eta$ are the usual scalar fields with masses $m_1$ and $m_2$,
respectively.

\noindent Interaction can be included in equation (\ref{FSF})
following the pattern (\ref{ODE2}), (\ref{FU'}) of previous section:
\begin{equation}\label{ISF}
{T}\equiv\frac{(\Box+m_1^2)(\Box+m_2^2)\phi}{(m_1^2-m_2^2)}-
U'\left(\frac{(\alpha+\beta)\Box+(\alpha m_2^2+\beta
m_1^2)}{m_2^2-m_1^2}\phi\right)=0\,.
\end{equation}
The common multiplier $m^2_2-m_1^2$ provides the correct dimension
of energy.

Here we consider $U$ that does not depend on derivatives of fields.
This allows us to simplify explicit formulas in this section. The
general expressions and conclusions, however,  hold true even if the
interaction depends on the derivatives of fields.

The corresponding energy-momentum tensor reads
\begin{equation}\label{TEM}
\Theta^{\mu}_{\nu}=\alpha
\Theta^{(1)}{}^{\mu}_{\nu}(\mathcal{Q}\phi)+\beta
\Theta^{(2)}{}^{\mu}_{\nu}(\mathcal{P}\phi)+\delta^\mu_\nu U
\left(\frac{(\alpha+\beta)\Box+(\alpha m_2^2+\beta
m_1^2)}{m_2^2-m_1^2}\phi\right)\,,
\end{equation}
where
$$\begin{array}{ll}\displaystyle
\Theta^{(1)}{}^{\mu}_{\nu}(\mathcal{Q}\phi)&=\displaystyle\partial^\mu\Big(\frac{\Box\phi+m_2^2\phi}{m_2^2-m_1^2}\Big)\partial_\nu\Big(\frac{\Box\phi+m_2^2\phi}{m_2^2-m_1^2}\Big)-
\\[3mm]&\displaystyle-\frac{1}{2}\delta^{\mu}_{\nu}\partial^\sigma\Big(\frac{\Box\phi+m_2^2\phi}{m_2^2-m_1^2}\Big)\partial_\sigma\Big(\frac{\Box\phi+m_2^2\phi}{m_2^2-m_1^2}\Big)+
\delta^{\mu}_{\nu}\frac{m_1^2}{2}\Big(\frac{\Box\phi+m_2^2\phi}{m_2^2-m_1^2}\Big)^2
\end{array}$$
and
$$\begin{array}{ll}\displaystyle
\Theta^{(2)}{}^{\mu}_{\nu}(\mathcal{P}\phi)&=\displaystyle\partial^\mu\Big(\frac{\Box\phi+m_1^2\phi}{m_1^2-m_2^2}\Big)\partial_\nu\Big(\frac{\Box\phi+m_1^2\phi}{m_1^2-m_2^2}\Big)-
\\[3mm]&\displaystyle-\frac{1}{2}\delta^{\mu}_{\nu}\partial^\sigma\Big(\frac{\Box\phi+m_1^2\phi}{m_1^2-m_2^2}\Big)\partial_\sigma\Big(\frac{\Box\phi+m_1^2\phi}{m_1^2-m_2^2}\Big)+
\delta^{\mu}_{\nu}\frac{m_2^2}{2}\Big(\frac{\Box\phi+m_1^2\phi}{m_1^2-m_2^2}\Big)^2\,
\end{array}$$
are the energies of scalar modes with masses $m_1$ and
$m_2$, and the last term in (\ref{TEM}) has the sense of interaction
energy.

The characteristic of the conserved energy-momentum tensor
(\ref{TEM}) reads
\begin{equation}\label{CharSF}
Q_\nu=\partial_\nu \left(\frac{(\alpha+\beta)\Box+(\alpha
m_2^2+\beta m_1^2)}{m_1^2-m_2^2}\phi\right)\,,\qquad
\partial_\mu \Theta^\mu_\nu=Q_\nu{T}\,.
\end{equation}
The Lagrange anchor, being constructed for equation (\ref{ISF}) by
the general recipe (\ref{V}), has the form
\begin{equation}\label{LASF}
V=\frac{1}{\alpha}\frac{\Box+m_1^2}{m_1^2-m_2^2}+\frac{1}{\beta}\frac{\Box+m_2^2}{m_1^2-m_2^2}+
\frac{1}{m_2^2-m_1^2}\frac{(\alpha+\beta)^2}{\alpha\beta}\frac{d^2U}{d\phi^2}\left(\frac{(\alpha+\beta)\Box+(\alpha
m_2^2+\beta m_1^2)}{m_2^2-m_1^2}\phi\right)\,.
\end{equation}
The Lagrange anchor maps characteristics to infinitesimal symmetry
transformations, see Appendix B. Applying the anchor (\ref{LASF}) to
the characteristic (\ref{CharSF}), we find
$$
\delta_\varepsilon\phi=\varepsilon^\mu V (Q_\mu)=-\varepsilon^\mu
\partial_\mu\phi-\frac{(\alpha+\beta)^2}{\alpha\beta}\frac{1}{m_1^2-m_2^2}\varepsilon^\mu\partial_\mu
T\,,
$$
where $T$ is the l.h.s. of the field equation (\ref{ISF}). The
symmetry transformation is a translation along the constant vector
$\varepsilon^\mu$, as it must be. The stable interaction vertices
correspond to $\alpha,\beta>0$ and depend on the second derivatives
of the scalar field through $\Box\phi$.

In Ref. \cite{AEBV} the higher derivative self-interactions of the
scalar field of the similar form are considered in cosmology as one
of the scenarios explaining inflation. With this regard, the
suggested stability control method, being based on the conservation
of the tensor (\ref{TEM}), can be relevant to cosmology where the
classical stability is an important selection principle for the
models.

Let us mention one more evidence of stability of scalar fields with
high derivatives. The instability of the theory is usually related
with the presence of  ``ghost states''. These states correspond to
the wrong sign of the pole in propagator. They are responsible for
the presence of negative norm states that represents a notorious
trouble of high derivative theories. Below we demonstrate that the
correct choice of the Lagrange anchor leads to the ghost-free
theory. The procedure of quantization of theories equipped with the
Lagrange anchor has been developed in the series of works
\cite{KazLS,LS1,LS2}. Here, we use the method based on the
generalized Schwinger-Dyson equation (A brief outline of the method
can be found in the Appendix A, for more systematic exposition see
\cite{LS1}). We find the generating functional of Green's functions
for the free higher-derivative scalar field with Lagrange anchor
(\ref{LASF}) and derive the propagator as the second variational
derivative of the generating functional of Green's functions.

For the free equations of motion (\ref{FSF}) and the
Lagrange anchor (\ref{LASF}), the Schwinger-Dyson equation reads
\begin{equation}\label{GSDE-SF}
\left[\frac{\delta S}{\delta \phi} (\widehat\phi)
-V(\bar\phi)\right]Z[\bar\phi]=0\,,
\end{equation}
where $\widehat\phi=i\hbar \delta/\delta\bar\phi$, $\bar\phi$ is the
source for the scalar field $\phi$, and $Z[\bar\phi]$ is the
generating functional of Green's functions. The solution to the
Schwinger-Dyson equation (\ref{GSDE-SF}) has the form
\begin{equation}\label{GSDE-SOL}
    Z[\bar\phi]=\exp\left[-\frac{i}{2\hbar}\int d^4x \bar{\phi}\left(\frac{1}{\alpha}\frac{1}{\Box+m_2^2}+
    \frac{1}{\beta}\frac{1}{\Box+m_1^2}\right)\bar\phi\right]\,.
\end{equation}
Taking the second variational derivative of (\ref{GSDE-SOL}) and setting
$\bar\phi=0$, we get the propagator
\begin{equation}\label{P-SF}
    G_2(x_1-x_2)=i\hbar\frac{\delta^2
    Z[\bar\phi]}{\delta\bar\phi(x_1)\delta\bar\phi(x_2)}\Big|_{\bar\phi=0}=\Big(\frac{1}{\alpha}\frac{1}{\Box+m_2^2}+
    \frac{1}{\beta}\frac{1}{\Box+m_1^2}\Big)\delta(x_1-x_2)\,.
\end{equation}
As one could expect, both the terms in (\ref{P-SF}) have the same
sign if $\alpha, \beta>0$. The canonical Lagrange anchor corresponds
to the choice $\alpha=-\beta=1$ that leads to the theory with
ghosts.

Let us note that the presence of derivatives in the Lagrange anchor
makes the ultraviolet behavior of the propagator  worse. Only the
canonical Lagrange anchor ($\alpha=-\beta$) provides the ultraviolet
asymptotic form $G_2\sim p^{-4}$ in the momentum representation. In
the case of positive definite energy, the propagator behaves like
the usual Feynman's propagator for the scalar field, $G_2\sim
p^{-2}$. As a result, the use of Lagrange anchor with derivatives
does not allow one to get simultaneously the positive definite
energy and improve the renormalization properties of the theory.
This can decrease the potential attractiveness of using
higher-derivative theories from the viewpoint of surpassing the
divergences in quantum theory.

As we have seen, at free level the higher derivative scalar field
model admits a two-parameter family of conserved energy-momentum
tensors. The interaction, being included by the recipe (\ref{ISF}),
explicitly involves these parameters. In the interacting model only
one conservation law survives by construction. The conserved tensor
(\ref{EMT}) has positive density $\Theta^0_{\phantom{0}0}$ once
$\alpha,\beta >0$, while the canonical energy (which is unbounded)
corresponds to $\alpha=-\beta=1$. So, the interaction with $\alpha,
\beta >0$ does not break stability, because the positive quantity
still conserves in this case. A similar phenomenon is seen when the
theory is quantized. If the Lagrange anchor is chosen with positive
parameters $\alpha,\beta$ the theory is stable, while the canonical
choice results in the ghosts.

\subsection{Podolsky's electrodynamics and its interaction with massive spin $1/2$}
The free Podolsky's electrodynamics is the theory of vector field
$\phi^\mu$ with action
\begin{equation}\label{PSED}
S=-\frac{1}{4}\int dx
\Big[(F_\phi)_{\mu\nu}(F_\phi)^{\mu\nu}-\frac{2}{m^2_p}\partial^\mu
(F_\phi)_{\mu\rho}\,\partial_\nu (F_\phi)^{\nu\rho}\Big]\,.
\end{equation}
Here,
$(F_\phi)_{\mu\nu}=\partial_\mu\phi_\nu-\partial_{\nu}\phi_\mu$ is
the field strength and $m_p>0$ is the parameter of theory having
the dimension of mass.

The equations of motion
$$
-\frac1 {m_p^2}\frac{\delta S}{\delta
\phi}\equiv\mathcal{P}\mathcal{Q}\phi=0\,
$$
have factorable structure (\ref{ODE2}), where the operators
$\mathcal{P}, \mathcal{Q}$ and $ \mathcal{F} $ read
\begin{equation}\label{PSPQ}
  \mathcal{P}=-\frac1{m_p^2}(\Box-\partial\partial\cdot)\,,\qquad
  \mathcal{Q}=\frac1{m_p^2}(\Box-\partial\partial\cdot+m_p^2)\,,\qquad
  \mathcal{F}=0 \ .
\end{equation}
Obviously $\mathcal{P}$ is the Maxwell operator, $\mathcal Q$ is the
Proca operator.

Being a factorable fourth-order  theory, the Podolsky
electrodynamics can be reduced to the second order by introducing
the variables $\xi$ and $\eta$ that absorb the second derivatives of
$\phi$ following the general recipe (\ref{map}):
$\xi=\mathcal{Q}\phi$, $\eta=\mathcal{P}\phi$. Then, the equivalent
second-order theory will be given by the Maxwell equations for $\xi$
and the Proca equations for $\eta$. The corresponding action has the
form
\begin{equation}\label{PSL}
S_1=-\frac{1}{4}\int dx \Big[\alpha
(F_\xi)_{\mu\nu}(F_\xi)^{\mu\nu}+
\beta\left((F_\eta)_{\mu\nu}(F_\eta)^{\mu\nu}-2m^2_p\eta^\nu\eta_\nu\right)\Big]\,
\end{equation}
with some constants $\alpha,\beta\neq0$. The Lagrangians
(\ref{PSED}) and (\ref{PSL}) enjoy the usual gauge symmetry
\begin{equation}\label{GSym1}
  \delta_\chi \phi_\mu=\partial_\mu\chi\,,\qquad\delta_\chi\xi_\mu=\partial_\mu\chi\,,\qquad \delta_\chi \eta_\mu=0\,.
\end{equation}

Let us first discuss the inclusion of interaction in the
$\xi\eta$-formalism, and then switch to the $\phi$-picture, where
the equations are of fourth order.\footnote{The second-order system
remains equivalent to the fourth order one once the interaction is
included following the pattern (\ref{ODE1}). If the interacting
second-order system is not factorable in the sense of (\ref{ODE1}),
it can be inequivalent to any fourth-order system.} Introduce the
Dirac field $\psi$ ($\widetilde{\psi}$ stands for the Dirac
conjugate spinor) minimally coupled to the vector field by adding
the following term to the action (\ref{PSL}):
\begin{equation}\label{PSLint}
  S'_1=S_1-\int dx U\,,\qquad
  U(\alpha\xi-\beta\eta,\psi,\widetilde\psi)=-\widetilde{\psi}
  (i\gamma^\mu(\partial_\mu-e(\alpha\xi-\beta\eta)_\mu)-m)\psi\,.
\end{equation}
The equations read
\begin{equation}\label{PSMaxPro}
\partial^\nu(F_\xi)_{\nu\mu}-j_\mu=0\,,
\qquad \partial^\nu(F_\eta)_{\nu\mu}+m_p^2\eta_\mu+j_\mu=0\,, \qquad
j_\mu=e\widetilde{\psi}\gamma^\mu\psi\,,
\end{equation}
\begin{equation}\label{PSD}
     (i\gamma^\mu(\partial_\mu-e(\alpha\xi-\beta\eta)_\mu)-m)\psi=0\,,\qquad
     \widetilde{\psi}(i\gamma^\mu(\overleftarrow{\partial}_\mu+e(\alpha\xi-\beta\eta)_\mu)+m)=0\,.
\end{equation}
The consistency of interaction implies that the gauge
transformations (\ref{GSym1}) are complemented by the standard
$U(1)$-transformation for the Dirac field
\begin{equation}\label{GSym2}
\delta_\chi\psi= -ie\alpha\chi\psi\,,\qquad
\delta_\chi\widetilde\psi=ie\alpha\chi\widetilde\psi.
\end{equation}
As is seen, the full theory (\ref{PSMaxPro}), (\ref{PSD}) describes
propagation of one vector field $\eta$ of mass $m_p$ and one
massless gauge field $\xi$, and both the vectors are minimally
coupled to the spinor field $\psi$.

If $\alpha,\beta>0$, the theory (\ref{PSL}) is (perturbatively)
stable. The energy-momentum tensor reads
\begin{equation}\label{EMTPS}\begin{array}{l}\displaystyle
\Theta^{\mu}_{\phantom{a}\nu}(\xi,\eta,\psi,\widetilde\psi)=\frac{\beta}{4}(
\delta^{\mu}_{\nu}(F_{\eta})^{\rho\sigma}(F_{\eta})_{\rho\sigma}-
4(F_{\eta})^{\mu\rho}(F_{\eta})_{\nu\rho}+ 4m_p^2 \eta^\mu
\eta_\nu-2m_p^2\delta^{\mu}_{\nu}\eta^\rho\eta_\rho)+\\[3mm]\qquad\displaystyle+
\frac{\alpha}{4}(\delta^{\mu}_{\nu}(F_{\xi})^{\rho\sigma}(F_{\xi})_{\rho\sigma}-
4(F_{\xi})^{\mu\rho}(F_{\xi})_{\nu\rho})+
\frac{i}{4}\widetilde{\psi}\Big[\gamma^\mu(\overrightarrow{\partial}_\nu+ie(\alpha\xi-\beta\eta)_\nu)+
\\[3mm]\qquad\displaystyle+\gamma_\nu(\overrightarrow{\partial}^\mu+ie(\alpha\xi-\beta\eta)^\mu)-
\gamma^\mu(\overleftarrow{\partial}_\nu-ie(\alpha\xi-\beta\eta)_\nu)-
\gamma_\nu(\overleftarrow{\partial}^\mu-ie(\alpha\xi-\beta\eta)^\mu)\Big]\psi\,.
\end{array}\end{equation}

Notice that the stable and unstable models describe different
physics. To demonstrate this fact, let us make the field
redefinition
\begin{equation}\label{xi-xi}
\xi\rightarrow\pm\frac{\xi}{\sqrt{|\alpha|}},\qquad\eta\rightarrow\pm\frac{\eta}{\sqrt{|\beta|}}\,
\end{equation}
in the action (\ref{PSLint}). Substituting (\ref{xi-xi}) into
(\ref{PSLint}), we get the standard action of theory describing the
minimal coupling of massive and massless vector fields with Dirac
field
\begin{equation}\label{S1prime}\begin{array}{ll}\displaystyle
S'_1=&\displaystyle-\frac{1}{4}\int dx\Big\{
\frac{\alpha}{|\alpha|}(F_\xi)_{\mu\nu}(F_\xi)^{\mu\nu}+
\frac{\beta}{|\beta|}\Big[
(F_\eta)_{\mu\nu}(F_\eta)^{\mu\nu}-2m^2_p\eta^\nu\eta_\nu\Big]-\\[3mm]
&\displaystyle-4
\widetilde{\psi}\Big(i\gamma^\mu(\partial_\mu-e(\pm\frac{\alpha}{|\alpha|}\sqrt{|\alpha|}\xi
\mp\frac{\beta}{|\beta|}\sqrt{|\beta|}\eta)_\mu)-m\Big)\psi\Big\}\,.
\end{array}\end{equation}
The parameters $\alpha$, $\beta$ define the intensity of this
coupling. Notice that, by construction, any model (\ref{S1prime})
with nonzero $\alpha,\beta$ remains equivalent to the Posdolsky
theory interacting with Dirac field. By this reason, any theory of
massive and massless vector fields minimally interacting with spinor
field has equivalent description in terms of the interacting
Podolsky's theory.

It is well known that in the theory of the form (\ref{S1prime}), two
fermions interact by means of massless ``photons'' producing the
Coulomb force and massive ``photons'' producing the Yukawa force. If
the theory is stable, both types of photons mediate the force of
repulsion between two particles of the same charge and the force of
attraction if the particles have opposite electric charges. In
contrast, the unstable theories (because of the ``wrong'' sign of
action of one (or both) photons in (\ref{S1prime})) describe the
interactions where one (or both) types of photons mediate the force
of attraction between two particles of the same charge and the force
of repulsion between particle and antiparticle. For example, in the
special case of $\alpha=-\beta=1$ that corresponds to the inclusion
of minimal interaction $\phi^\mu j_\mu$ into the original Lagrangian
(\ref{PSPQ}), the Coulomb and Yukawa contributions to the
interaction energy are equal by intensity but \textit{must} be
different by sign. This fact was first noticed by Podolsky in
\cite{Podolsky1942} and it was turned out that this sign cannot be
controlled within the Lagrangian formalism. It was long believed
that the phenomenon of subtraction two forces is the strong side of
the theory, because it allows one to make better the short-distance
behavior of Green's functions. Now we see that the minimal
interaction of Podolsky theory with Dirac field is incompatible with
the stability condition. The stable interactions with
$\alpha,\beta>0$ correspond to non-minimal and non-Lagrangian
interaction vertices in the Podolsky theory. Below, we explain that
the stability of the theory can be controlled immediately it terms
of fourth-order equations with any $\alpha,\beta$ even though they
are not necessarily Lagrangian.

In the $\phi$-representation, that corresponds to the original
fourth-order formalism, the equations of nonlinear theory
(\ref{PSMaxPro}), (\ref{PSD}) read

\begin{equation}\label{TPS}\begin{array}{l}
\displaystyle
(T_\phi)_\mu\equiv\Big(\frac{1}{m^2_p}\Box+1\Big)\partial^\nu
(F_\phi)_{\mu\nu}-j_\mu\,, \\[4mm]
\displaystyle T_{\widetilde{\psi}}\equiv
\Big\{i\gamma^\mu\Big(\partial_\mu-e\alpha\phi_\mu-e\frac{\alpha+\beta}{
m^2_p}\partial^\nu (F_\phi)_{\nu\mu}\Big)-m\Big\}\psi\,,\\[4mm]
\displaystyle T_{\psi}\equiv
\widetilde\psi\Big\{i\gamma^\mu\Big(-\overleftarrow{\partial}_\mu-e\alpha\phi_\mu-e\frac{\alpha+\beta}{
m^2_p}\partial^\nu (F_\phi)_{\nu\mu}\Big)-m\Big\}\,.
\end{array}\end{equation}
The equations (\ref{TPS}) are invariant under the usual gauge
transformations (\ref{GSym1}), (\ref{GSym2}).

In the $\phi$-representation the energy-momentum tensor
(\ref{EMTPS}) takes the form
\begin{equation}\label{EMTPS2}\begin{array}{l}
\displaystyle\Theta^{\mu}_{\phantom{a}\nu}(\phi,\psi,\widetilde\psi)=
\frac{\alpha+\beta}{4m^4_p}\Big[\delta^\mu_{\nu}(\Box
F_\phi)^{\rho\sigma}(\Box F_\phi)_{\rho\sigma}-4(\Box
F_\phi)^{\mu\rho}(\Box
F_\phi)_{\nu\rho}\Big]+\\[3mm]\displaystyle+\frac{\alpha}{2m_p^2}\Big[
\delta^\mu_{\nu} (F_\phi)^{\rho\sigma}(\Box
F_\phi)_{\rho\sigma}-2(F_\phi)^{\mu\rho}(\Box
F_\phi)_{\nu\rho}-2(F_\phi)_{\nu\rho}(\Box
F_\phi)^{\mu\rho}\Big]+\\[3mm]\displaystyle+\frac{\beta}{2m_p^2}\Big[2\partial_\rho
(F_\phi)^{\rho\mu}\partial^\sigma
(F_\phi)_{\sigma\nu}-\delta^{\mu}_{\phantom{\mu}\nu}\partial_\rho
(F_\phi)^{\rho\tau}\partial^\sigma (F_\phi)_{\sigma\tau}\Big]+
\frac{1}{4}\delta^\mu_{\nu}(F_\phi)^{\rho\sigma}(F_\phi)_{\rho\sigma}-
(F_\phi)^{\mu\rho}(F_{\phi})_{\nu\rho}+\\[3mm]\displaystyle+
\frac{i}{4}\widetilde{\psi}\Big[\gamma^\mu(\overrightarrow{\partial}_\nu+ieb_\nu)
+\gamma_\nu(\overrightarrow{\partial}^\mu+ieb^\mu)-
\gamma^\mu(\overleftarrow{\partial}_\nu-ieb_\nu)-
\gamma_\nu(\overleftarrow{\partial}^\mu-ieb^\mu)\Big]\psi\,,
\end{array}
\end{equation}
where
$$
b_\mu=\alpha\phi_\mu+\frac{\alpha+\beta}{ m^2_p}\partial^\nu
(F_\phi)_{\nu\mu}\,.
$$
In the limit of free Lagrangian theory ($\alpha=-\beta=1,\psi=0)$
this conserved tensor reduces to the standard energy-momentum tensor
of the Podolsky theory \cite{Podolsky1942} as one could expect.

The tensor (\ref{EMTPS2}) conserves,
\begin{equation}\label{divEMT}
\partial_\mu\Theta^{\mu}_{\phantom{a}\nu}=
\mathcal{}(Q_\phi)_\nu^\mu (T_\phi)_\mu+T_{\psi}(Q_\psi)_\nu
+(Q_{\widetilde{\psi}})_\nu T_{\widetilde{\psi}}\,,
\end{equation}
and the respective characteristic reads \footnote{This equality is
understood modulo equivalence. The Lagrange anchor maps equivalent
characteristics to equivalent symmetries. See for details Appendix
\ref{AppB} and \cite{KapLS2010,KapLS2011JHEP}.}
\begin{equation}\label{EMTPSChar}
Q_\nu=((Q_\phi)^\mu_\nu,(Q_\psi)_\nu,(Q_{\widetilde\psi})_\nu)=(-\partial_\nu b^\mu,
-\partial_\nu\psi,-\partial_\nu\widetilde\psi)\,.
\end{equation}
The Lagrange anchor (\ref{anchor}) for factorable systems is
constructed by the general recipe (\ref{V}). Following this pattern,
we arrive at the Lagrange anchor $V$, whose action on the general
characteristic $Q$ reads
\begin{equation}\label{VQPS}\begin{array}{l}
\displaystyle V(Q)\equiv\left(V^\mu_\phi(Q),V_{\bar{\psi}}(Q),V_\psi(Q)\right)=\\[3mm]
\displaystyle\qquad=\Big(\Big[\Big(\frac{1}{\alpha}+\Big(\frac{1}{\alpha}+
\frac{1}{\beta}\Big)\frac{\Box-\partial\partial\cdot}{m_p^2}\Big)Q\Big]^\mu+
\frac{1}{m_p^2}\frac{(\alpha+\beta)^2}{\alpha\beta}\Big[e\bar{\psi}\gamma^\mu
Q_{\bar{\psi}}+eQ_{\psi}\gamma^\mu\psi\Big], \ Q_\psi, \
Q_{\bar{\psi}}\Big)\,.
\end{array}\end{equation}
Substituting (\ref{EMTPSChar}) into (\ref{VQPS}), we find the
following symmetry transformation corresponding to the
characteristic:
\begin{equation}\label{deltaphiPS}
(\delta_\varepsilon\phi^\mu,\delta_\varepsilon\psi,\delta_\varepsilon\widetilde\psi)=
\varepsilon^\nu
V(Q_\nu)=(-\varepsilon^\nu\partial_\nu\phi^\mu-\frac{1}{m^2_p}\frac{(\alpha+\beta)^2}{\alpha\beta}
\varepsilon^\nu\partial_\nu
(T_\phi)^\mu,-\varepsilon^\nu\partial_\nu\psi,-\varepsilon^\nu\partial_\nu\widetilde\psi)\,
.
\end{equation}
This means that the Lagrange anchor connects the conservation of the
tensor (\ref{EMTPS2}) with translation invariance of the
fourth-order equations (\ref{TPS}). Once $\alpha,\beta$ are
positive, the tensor satisfies the condition
$\Theta^0_{\phantom{0}0}>0$, and the theory is stable. The
corresponding positive, conserved, non-canonical energy-momentum
tensor is connected to the translation invariance by the
non-canonical Lagrange anchor (\ref{VQPS}).

If the fourth-order equations (\ref{TPS}) were quantized with the
corresponding Lagrange anchor with $\alpha >0, \beta > 0$ along the
lines of previous section, we would arrive at the stable quantum
theory precisely corresponding to the quantization of the
second-order Lagrangian (\ref{PSL}), (\ref{PSLint}). If the
fourth-order theory is considered with unstable vertices
corresponding to the opposite signs of $\alpha$ and $\beta$ in the
Lagrange anchor, the theory will be classically unstable, and it's
quantization will correspond to the standard Feynman rules for the
Podolsky Lagrangian with minimal coupling to the Dirac field. The
quantum instability problem is well known for the couplings of this
type, see e.g. \cite{Bufalo2011, Bufalo2012, Bufalo2013} and
references therein.

In this section, we have studied the stability proceeding from the
fact that the free higher-derivative electrodynamics by Podolsky has
the factorable structure of equations. Because of that, it admits a
bounded conserved energy--momentum tensor, besides the unbounded
canonical one. The conservation of the bounded tensor ensures
classical stability irrespectively to unboundness of the canonical
tensor. Then, we considered not necessarily minimal inclusion of
interactions with the massive spin $1/2$ field such that the bounded
tensor, being deformed by the interaction (\ref{EMTPS}), still keeps
conserving. The nonlinear higher-derivative theory is both
classically and quantum mechanically equivalent to the theory of one
massless and one massive vector fields both coupled with the Dirac
field. Studying these auxiliary second-order formulations, we showed
that the minimal coupling of Podolsky's theory breaks stability of
the free theory, while the non-minimal interactions (\ref{TPS}) keep
the dynamics stable.

\section{Conclusion}
In this paper we study the higher-derivative dynamics proceeding
from the idea that the stability can be ensured by the existence of
any bounded conserved quantity even if it is different from the
canonical energy. We have focused at the special class of factorable
higher-derivative systems whose equations (\ref{ODE2}) include the
linear term $\mathcal{PQ}\phi$ and the nonlinearity
$\mathcal{F}(\mathcal{P}\phi ,\mathcal{Q}\phi )$. By making use of
factorization, we can construct the conserved quantity that might be
positive both in linear model and with a variety of interactions
$\mathcal{F}$, while the canonical energy is not positive definite
for the system already in the linear approximation. The conservation
of this positive quantity is by construction connected to the
translation invariance, so it can be viewed as an alternative
definition of energy for the higher-derivative systems. As we have
demonstrated, the classical stability can be promoted to the quantum
level. This class of higher-derivative systems is wide enough to
accommodate the models of interest for physics, as is seen from the
examples of Section 4. However, the factorable structure of
equations seems us to be rather a technical tool than a genuine
restriction for the dynamics related to stability. In any case, we
see that higher-derivative systems can have stable classical and
quantum dynamics with non-trivial interactions irrespectively to the
fact that the canonical energy is unbounded.

\vspace{0.2 cm}

\noindent \textbf{Acknowledgements.} The authors thank I.V. Tyutin
for discussions. The work is partially supported by the Tomsk State
University Competitiveness Improvement Program and the RFBR grant
13-02-00551. A.Sh. appreciates the financial support from Dynasty
Foundation.

\appendix
\numberwithin{equation}{section}

\section{The Lagrange anchor}\label{AppA}
The appendix provides an elementary introduction to the concept of
the Lagrange anchor. A more systematic and rigorous exposition of
the subject can be found in \cite{KazLS, LS1, LS2,KapLS2010}.

In the quantum field theory one usually studies the path integrals
of the form
\begin{equation}\label{PI}
    \langle \mathcal{O}\rangle =\int [d\varphi]
    \,\mathcal{O}[\varphi]\,e^{\frac i{\hbar}S[\varphi]}\,.
\end{equation}
After normalization, this integral defines the quantum average of an
observable $\mathcal{O}[\varphi]$ in the theory with action
$S[\varphi]$. Here $\varphi=\{\varphi^i\}$ is a collection of fields
on a space-time manifold $M$. It is believed that evaluating the
path integrals for various observables $\mathcal{O}$, one can
extract all physically relevant information about the quantum
dynamics of the model.

The  functional $\Psi [\varphi]=e^{\frac i{\hbar}S[\varphi]}$,
weighting the contribution of a particular field configuration
$\varphi$ to the quantum average, is known as the Feynman
probability amplitude on the configuration space of fields. This
amplitude can be defined as a unique (up to a normalization factor)
solution to the Schwinger-Dyson (SD) equation\footnote{Here we use
the condensed index notation \cite{deWitt}, so that the partial
derivatives with respect to fields should be understood as
variational ones and summation over the repeating indices includes
integration over $M$.}
\begin{equation}\label{SDf}
   \left (\frac{\partial S}{\partial\varphi^i}+i\hbar\frac{\partial}{\partial
    \varphi^i}\right) \Psi[\varphi]=0\,.
\end{equation}
Performing the Fourier transform from the fields $\varphi$ to their
sources $\bar\varphi$, we can bring (\ref{SDf}) to a more familiar
form
\begin{equation}\label{SD}
\left( \frac{\partial
S}{\partial\varphi^i}(\widehat{\varphi})-\bar\varphi_i\right)Z[\bar\varphi]=0\,,\qquad
\widehat{\varphi}{}^i\equiv i\hbar\frac{\partial}{\partial
\bar\varphi_i}\,,
\end{equation}
where
\begin{equation}\label{Z}
    Z[\bar\varphi]=\int [d\varphi] e^{\frac{i}{\hbar}(S[\varphi]-\bar\varphi\varphi)}
\end{equation}
is the generating functional of Green's functions.

The following observations provide guidelines for the generalization
of the Schwinger-Dyson equation to non-Lagrangian field theory, and
finding alternatives for the Lagrangian models.

\vspace{2mm}

    $(i)$ Although the Feynman probability amplitude involves an action functional,
    the SD  equations (\ref{SDf})
    contain solely the classical field equations, not the action as such.

\vspace{2mm}

    $(ii)$ In the classical limit $\hbar\rightarrow 0$, the second
    term in the SD equation (\ref{SDf}) vanishes, and the
    Feynman probability amplitude $\Psi$ turns into the Dirac distribution supported
    at the classical solutions to the field equations. Formally, $\Psi[\varphi]|_{\hbar\rightarrow 0} \sim
    \delta[\partial_i
    S]$ and one can think of the last expression as classical probability amplitude.

\vspace{2mm}

    $(iii)$ It is quite natural to treat the sources $\bar\varphi$ as the momenta
    canonically conjugate  to the fields $\varphi$,
    so that the only non-vanishing Poisson brackets are
    $\{\varphi^i,\bar\varphi_j\}=\delta^i_j$.
    Then, one can regard the SD operators
\begin{equation}\label{SDOP}
\frac{\partial S}{\partial\varphi^i}+i\hbar\frac{\partial}{\partial
    \varphi^i}
\end{equation}
involved in (\ref{SDf})  as resulting from the canonical
quantization of the first class constraints
\begin{equation}\label{CLA}
    \partial_i S[\varphi]-\bar{\varphi}_i\approx 0
    \end{equation}
    on the phase space of fields and sources. Upon this
    interpretation, the Feynman probability amplitude describes  a unique physical state
    of a first-class constrained theory.
    This state is unique because  the ``number'' of the first class constraints (\ref{CLA}) is equal  to the
    ``dimension'' of the configuration space of fields. Quantizing the constrained system (\ref{CLA}) in
    the   momentum representation, one arrives at the SD equation (\ref{SD}) for
    the partition function $Z[\bar\varphi]$.

\vspace{2mm}

The above interpretation of the SD equations as operator first class
constraints on a physical wave-function suggests a direct way to
their generalization. Consider a set of field equations
\begin{equation}\label{T-eq}
T_a(\varphi)=0\,,
\end{equation}
which do not necessarily follow from the variational principle. In
this case, the (discrete parts of) superindices $a$ and $i$ may run
over different sets. Proceeding from the heuristic arguments above,
we can take the following ansatz for the $\varphi
\bar\varphi$-symbols of the Schwinger-Dyson operators:
\begin{equation}\label{TT}
\mathcal{T}_a=T_a(\varphi)-V_a^i(\varphi)\bar\varphi_i+
O(\bar\varphi^2)\,.
\end{equation}
The symbols are defined as formal power series in  sources
$\bar\varphi$ with leading terms being the classical equations of
motion. Requiring the Hamiltonian constraints $\mathcal{T}_a\approx
0$ to be first class, i.e.,
\begin{equation}\label{inv}
    \{\mathcal{T}_a, \mathcal{T}_b\}=U_{ab}^c \mathcal{T}_c \,,\qquad
    U_{ab}^c(\varphi,\bar\varphi)=C^c_{ab}(\varphi)+ O(\bar\varphi)\,,
\end{equation}
we obtain an infinite set of relations on the expansion coefficients
of $\mathcal{T}_a$ in the powers of sources. In particular,
verifying the involution relations (\ref{inv}) up to zero order in
$\bar\varphi$, we find
\begin{equation}\label{anchor}
    V_a^i\partial_iT_b-V_b^i\partial_iT_a=C_{ab}^c T_c\,
\end{equation}
for some structure functions $C^c_{ab}(\varphi)$. The value
$V_a^i(\varphi)$ defined by (\ref{anchor}) is called the
\textit{Lagrange anchor}.

For variational field equations, $T_a=\partial_i S$, one can set the
Lagrange anchor to be the unit matrix $V_a^i=\delta^i_a$. This
choice results in the standard Schwinger-Dyson operators
(\ref{SDOP}) obeying the abelian involution relations. For this
reason we refer to $V_a^i=\delta_a^i$ as the \textit{canonical
Lagrange anchor} of the Lagrangian dynamics.  Generally, the
Lagrange anchor may be field-dependent and/or noninvertible. If the
Lagrange anchor is invertible (in which case the number of equations
must coincide with the number of fields), then the operator $V^{-1}$
plays the role of integrating multiplier in the inverse problem of
calculus of variations. So, the existence of the invertible Lagrange
anchor is equivalent to the existence of action. The other extreme
choice, $V=0$, is always possible and corresponds to the classical
probability amplitude $\Psi[\varphi]\sim\delta[T_a(\varphi)]$
supported at the classical solutions. Any nontrivial Lagrange
anchor, be it invertible or not, yields a fuzzy partition function
describing nontrivial quantum fluctuations in the directions spanned
by the vector fields $V_a = V_a^i\partial_i$.

In the non-Lagrangian case, the constraints (\ref{TT}) are not
generally the whole story. The point is that the number of
(independent) field equations may happen to be less than the number
of fields. In this case, the field equations (\ref{T-eq})  do not
specify a unique solution with prescribed boundary conditions or,
stated differently, the system enjoys a gauge symmetry  generated by
an on-shell integrable vector distribution
$R_\alpha=R_\alpha^i(\varphi)\partial_i$ such that
\begin{equation}\label{}
R_\alpha^i\partial_i T_a=U_{\alpha a}^b T_b \,,\qquad [R_\alpha,
R_\beta]=U_{\alpha\beta}^\gamma R_\gamma + T_a
U_{\alpha\beta}^{ai}\partial_i
\end{equation}
for some structure functions $U^b_{\alpha a}(\varphi)$ and
$U_{\alpha\beta}^{ai}(\varphi)$.
 To take the gauge invariance into account at the quantum level, one has to
 impose additional first class constraints on the fields and sources. Namely,
\begin{equation}\label{R}
    \mathcal{R}_\alpha =R_\alpha^i(\varphi)\bar\varphi_i+O(\bar\varphi^2)\approx 0\,.
\end{equation}
The leading terms of these constraints coincide with the  $\varphi
\bar\varphi$-symbols of the gauge symmetry generators and the higher
orders in $\bar\varphi$ are determined from the requirement the
Hamiltonian constraints $\mathbb{T}_I=(\mathcal{T}_a,
\mathcal{R}_\alpha)$ to be in involution\footnote{For a Lagrangian
gauge theory we have $\mathcal{T}_i=\partial_iS-\bar\varphi_i$ and
$\mathcal{R}_\alpha =-R^i_\alpha \mathcal{T}_i=R_\alpha^i
\bar\varphi_i$. In this case, one may omit the ``gauge'' constraints
$\mathcal{R}_\alpha\approx 0$ as they are given by linear
combinations of the ``dynamical'' constraints $\mathcal{T}_i\approx
0$.}. With all the gauge symmetries included, the constraint surface
$\mathbb{T}_I\approx 0$ is proved to be a Lagrangian submanifold in
the phase space of fields and sources and the gauge invariant
probability amplitude is defined as a unique solution to the
\textit{generalized SD equation}
\begin{equation}\label{SDE}
    {\widehat{\mathbb{T}}}{}_I\Psi=0\,.
\end{equation}
The last formula is just the definition of a physical state in the
Dirac quantization method of constrained dynamics. A systematic
presentation of the generalized SD equation can be found in Refs.
\cite{KazLS,LS1,LS2}.

In what follows we will refer to the first class constraints
$\mathbb{T}_I \approx 0$ as the \textit{Schwinger-Dyson extension}
of the original equations of motion (\ref{T-eq}). Notice that the
defining relations (\ref{anchor}) for the Lagrange anchor together
with the ``boundary conditions'' (\ref{TT}) and (\ref{R}) do not
specify a unique SD extension for a given system of field equations.
One part of the  ambiguity is related to the canonical
transformations in the phase space of fields and sources. If the
generator $G$ of  a canonical transform is at least quadratic in
sources,
\begin{equation}\label{}
G=\frac12 G^{ij}(\varphi)\bar\varphi_i\bar\varphi_j +
O(\bar\varphi^3)\,,
\end{equation}
then the transformed constraints
\begin{equation}\label{G-tr}
\begin{array}{l}
    \mathcal{T}'_a=e^{\{G,\,\,\cdot\,\,\}}\mathcal{T}_a=T_a-(V_a^i
    +G^{ij}\partial_j T_a)\bar\varphi_i + O(\bar\varphi^2)\,,\\[5mm]
    \mathcal{R}'_\alpha =
    e^{\{G,\,\,\cdot\,\,\}}\mathcal{R}_\alpha=R_\alpha^i\bar\varphi_i+O(\bar\varphi^2)\end{array}
\end{equation}
are in involution and start  with the same equations of motion and
gauge symmetry generators. Another ambiguity stems from changing the
basis of the constraints:
\begin{equation}\label{U-tr}
\begin{array}{l}
    \mathcal{T}''_a=U^b_a\mathcal{T}_b +U_a^\alpha \mathcal{R}_\alpha = {T}_a -(V_a^i+ A_a^{bi}T_b+B_a^\alpha
    R_\alpha^i)\bar\varphi_i+O(\bar\varphi^2)\,,\\[5mm]
    \mathcal{R}''_\alpha=U_\alpha^\beta
    \mathcal{R}_\beta+U_\alpha^a
    \mathcal{T}_a=R_\alpha^i\bar\varphi+O(\bar\varphi^2)\,,
    \end{array}
\end{equation}
where
\begin{equation}\label{}
\begin{array}{ll}
    U^b_a=\delta_a^b-A_a^{bi}\bar\varphi_i+O(\bar\varphi^2)\,,&\qquad
    U_a^\alpha=-B_a^\alpha + O(\bar\varphi)\,,\\[5mm]
    U_\alpha^\beta=\delta_\alpha^\beta+O(\bar\varphi)\,,&\qquad
    U_\alpha^a=O(\bar\varphi)\,.
    \end{array}
\end{equation}
Combining (\ref{G-tr}) with (\ref{U-tr}), we see that the Lagrange
anchor is defined modulo the equivalence relation
\begin{equation}\label{eqw}
V_a^i \sim V_a^i+ T_b A_a^{bi}+B_a^\alpha R_\alpha^i
+G^{ij}\partial_jT_a\,.
\end{equation}
The equivalent Lagrange anchors lead to essentially the same quantum
theory. We say that a Lagrange  anchor is trivial if it is
equivalent to the zero one.

\section{The generalized Noether theorem for (non-)Lagrangian theories }
\label{AppB}

The concept of Lagrange anchor allows one not only to quantize a
given (non-)Lagrangian theory, but also establish a correspondence
between its  symmetries and conservation laws. Unlike the classical
Noether's theorem this correspondence  is far from being canonical
and strongly depend on the choice of a particular Lagrange anchor.
Let us  recall  some basic definitions and constructions from
\cite{KapLS2010}.

An infinitesimal transformation of fields
$\delta_\varepsilon\varphi^i=\varepsilon Z^i(\varphi)$ is called a
symmetry of the equations of motion (\ref{T-eq}) if it  preserves
the mass shell, that is,
\begin{equation}\label{AppSym}
    \delta_\varepsilon T_a|_{T=0}=0\,,
\end{equation}
where $\varepsilon$ is a constant parameter. Two global symmetries
are considered as equivalent if they differ on shell by a gauge
symmetry transformation. In particular, adding to the generator
$Z^i$ any terms proportional to the equations of motion and their
differential consequences does not change its equivalence class.

A vector field $j^\mu(x,\varphi^i,\partial_\mu\varphi^i,\ldots)$ on
$M$ is called a conserved current if its divergence vanishes on
shell. For the regular field equations $T_a(\varphi,
\partial\varphi, \partial^{(2)}\varphi, \ldots \partial^{(k)}\varphi )=0$ this means the equality
\begin{equation}\label{dj}
 \partial_\mu j^\mu=
 \sum_{q=0}^p Q^{a,\mu_1\ldots\mu_q}
 (x,\varphi^i(x),\partial_{\mu}\varphi^i(x),\ldots)\partial_{\mu_1}\ldots\partial_{\mu_q}T_a\,\equiv Q^aT_a.
\end{equation}
The  differential operator  $Q$ is called the characteristic of the
conserved current $j$. Two conserved currents  $j$ and $j'$ are said
to be  equivalent if $j^{\mu}-j'^{\mu}=\partial_{\nu}i^{\nu\mu}$ $
(\mathrm{mod\phantom{a}} T_a)$ for some bivector
$i^{\mu\nu}=-i^{\nu\mu}$. Clearly, the equivalent conserved currents
lead to the same conserved charge.  By definition, two
characteristics $Q$ and $Q'$ are equivalent if they correspond to
equivalent currents. This equivalence  allows one to simplify the
form of characteristics. One can see that in each equivalence class
of $j$ there is a representative with $Q$ being a zero order
differential operator $Q^a$. For such a representative equation
(\ref{dj}) can be written as
\begin{equation}\label{TP}
Q^a T_a=\int_M \partial_\mu j^\mu\,.
\end{equation}
Here $a$ is understood as a condensed index, so that the sum on the
left implies integration over $M$. As is well known there is a
one-to-one correspondence between the equivalence classes of
conserved currents and characteristics \cite{KapLS2010}.

Given a  Lagrange anchor, one can assign to any characteristic $Q$ a
variational vector field $V(Q)=Q^a V_a^i \partial_i$. The main
observation made in \cite{KapLS2010} was that $V(Q)$ generates a
symmetry of the field equations (\ref{T-eq}):
\begin{equation}\label{deltaT}
\delta_\varepsilon \varphi^i=\varepsilon V^i(Q)\,, \qquad
\delta_\varepsilon T_a=\varepsilon Q^b V_b^i\partial_i
T_a=\varepsilon (-\partial_i Q^b
V^i_a+Q^cC_{ac}^b)T_b\,,\end{equation} with $\varepsilon$ being an
infinitesimal constant parameter. These relations  follow
immediately from the definitions (\ref{anchor}), (\ref{TP}) and  the
obvious identity $\partial_i(Q^a T_a)\equiv0$.

Recall that according to Noether's first theorem \cite{K-S} any
global symmetry $\delta\varphi^i=\varepsilon Q^i$ of the action
functional $S[\varphi]$ gives rise to the conserved current $j$ with
characteristic $Q^i$:
\begin{equation}\label{Ntheor}
\delta_\varepsilon S=0 \qquad \Leftrightarrow\qquad \int_M
\partial_\mu j^\mu= Q^i\frac{\delta S}{\delta\varphi^i}\,.
\end{equation}
Since a symmetry of the action is also a symmetry of the equations
of motion, one can regard the Noether correspondence (\ref{Ntheor})
as a particular case of the general relation (\ref{deltaT}), where
$V$ is taken to be the canonical Lagrange anchor $V=1$. From this
perspective, the assignment
\begin{equation}\label{NT} Q^a\mapsto
Z^i=Q^aV_a^i
\end{equation}
can be viewed as a natural extension of the first Noether's theorem
to the case of non-Lagrangian theories.

Let us stress that the correspondence (\ref{NT}) between the
Lagrange anchors and characteristics on one side and the symmetries
on the other is far from being a bijection: One and the same
symmetry $Z$ can be represented by different pairs $(Q, V)$. This
allows one to assign  different conserved currents to a given
symmetry by making use of different Lagrange anchors. In particular,
a Lagrangian system may have several conserved currents associated
with time translation if one admits non-canonical Lagrange anchors.
In this paper, we use this fact to construct a positive definite
energy for some high-derivative theories.

\section{Lagrange anchors for linear systems}\label{AppC}
Here we will illustrate the general notion of a Lagrange anchor by
the example of linear systems of partial differential equations with
constant coefficients. These have the form
 \begin{equation}\label{TPF}
 T(\partial)\phi=0\,,
 \end{equation}
  where $T=T(\partial)$ is a matrix differential operators and
  $\varphi$
  is the unknown multi-component  function on $M$.
  For simplicity we will assume that the matrix $T$ is square,
  so that the number of equations coincides with the number of fields $\varphi$.
  The Klein-Gordon, Maxwell and Dirac equations are all of this type.
  In this class of equations, $T(\partial )$ is often called the \textit{wave operator}.  The necessary and sufficient condition for the equations (\ref{TPF}) to come from the least action principle is the formal self-adjointness  of the wave operator, i.e.,
 \begin{equation}
 T^\ast=T\,,
 \end{equation}
 where $T^\ast(\partial)=T^t(-\partial)$.

Given a system of free field equations (\ref{TPF}), it is quite
natural to look for the Lagrange anchors being field-independent
differential operators $V=V(\partial)$ such that satisfy the
relation (\ref{anchor}). Then the Swinger-Dyson extension (\ref{TT})
of the field equations (\ref{TPF}) is given by
\begin{equation}
T(\partial)\varphi +V(\partial)\bar\varphi\approx 0\,.
\end{equation}
As was explained in Appendix A, the last expression should be
understood as a set of first class constraints on the phase space of
fields and sources. Linearity in the phase space variables implies
that these constraints are of the first class iff they pairwise
commute to each other. Then the defining condition for the Lagrange
anchor (\ref{anchor}) takes the simple form
\begin{equation}\label{TV=VT}
TV=V^\ast T^\ast\ .\end{equation} If both the Lagrange anchor and
the wave operator are (anti-)self-adjoint, $T^\ast=\pm T$ and
$V^\ast=\pm V$, then (\ref{TV=VT}) reduces to the commutativity
condition
\begin{equation}\label{TV}
[T,V]=0\,.
\end{equation}
We see that the problem of finding the Lagrange anchors for a system
of free field equations (\ref{TPF}) reduces to the issue of finding
the matrix $V(\partial)$ that commutes with the given matrix
$T(\partial )$ of the wave operator. As the entries of both matrices
are polynomials in commuting $\partial$'s, it is essentially a
problem of linear algebra over the ring polynomials. This problem
admits, in principle, a systematic solution by means of appropriate
algebraic techniques \cite{Eisen}, most of which exploit the idea of
Gr\"obner's bases. Particular solutions of physical interest can
also be found from more elementary considerations\footnote{From the
viewpoint of algebra, the problem of identifying the local gauge
symmetries for a given system of free field equations is similar to
the problem finding the Lagrange anchor for the system. The
difference is that the gauge generators $R(\partial)$ span the
kernel of the matrix $T(\partial)$, while the anchor $V(\partial)$
satisfies equation (\ref{TV}). The general algebraic techniques for
solving the equations $T(\partial ) R(\partial)=0$ can be found in
Section 4 of Ref. \cite{FLS}. Here, we do not develop the similar
techniques for the anchor, though it could be done along the same
lines.}. In relativistic field theory, for example, the general
structure of the Lagrange anchor is strongly constrained by symmetry
requirements, so that one can try some natural Lorentz-invariant
ansatz for $V(\partial)$. If the matrix operator $T$ is
(anti-)self-adjoint and diagonal, one can then always choose $V$ to
be an arbitrary operator of the same type, because the diagonal
matrix differential operators with constant coefficients obviously
commute.

Another typical situation when one can easily construct a particular
solutions to (\ref{TV=VT}) is a factorable wave operator. In that
case $T=\mathcal{P}\mathcal{Q}$, where  $\mathcal{P}$ and
$\mathcal{Q}$ are  commuting, formally self-adjoint operators. Then
we can choose
\begin{equation}\label{VPQ}
V=\rho \mathcal{Q}+\sigma \mathcal{P}\,.
\end{equation}
Condition (\ref{TV}) is obviously satisfied for any constants $\rho$
and $\sigma$ and we get a $2$-parameter family of the Lagrange
anchors. A particular example of this construction is given by the
Pais-Uhlenbeck oscillator (\ref{LEPU}), where the linear combination
(\ref{VPQ}) takes the form
\begin{equation}\label{App2VPU}
    V=\frac{\rho}{\omega_2^2-\omega_1^2}\left(\frac{d^2\phantom{t}}{dt^2}+\omega_2^2\right)+
    \frac{\sigma}{\omega_1^2-\omega_2^2}\left(\frac{d^2\phantom{t}}{dt^2}+\omega_1^2\right)\,.
\end{equation}
In this case, not only do the operators $\mathcal{P}$ and
$\mathcal{Q}$ provide a multiplicative decomposition of the wave
operator (\ref{LEPU}), but they also define an additive
decomposition of the canonical Lagrange anchor,
$\mathcal{P}+\mathcal{Q}=1$.

In a general way, the higher the order of differential equations,
the greater number of inequivalent Lagrange anchors  they admit. Let
us illustrate this thesis by an ordinary differential equation of
the form
\begin{equation}\label{dn}
\left(\frac{d^{2n}}{dt^{2n}}+a_1\frac{d^{2(n-1)}}{dt^{2(n-1)}}+\cdots+a_n\right)\varphi=0\,.
\end{equation}
Once the wave operator is formally self-adjoint, the equation is
Lagrangian. From the above discussion it appears that any
differential operator $V=V^\ast$ with constant coefficients can
serve as a Lagrange anchor for (\ref{dn}). Most of Lagrange anchors
are  equivalent to each other. Indeed, due to the third term
\footnote{Since the equation we consider is not gauge invariant and
the anchors are field independent, the first two terms in
(\ref{eqw}) appear to be irrelevant.} in the equivalence relation
(\ref{eqw}) one can remove from $V$ all the derivatives of order
$\geq 2n$. The equivalence classes of Lagrange anchors (with
constant coefficients) are thus described by the $n$-parameter
family of differential operators
$$
V=v_1\frac{d^{2(n-1)}}{dt^{2(n-1)}}+v_1\frac{d^{2(n-2)}}{dt^{2(n-2)}}+\cdots+v_n\,.
$$
For $n=1$ (the case of the second-order Lagrangian equations) the
space of Lagrange anchors is one-dimensional and is generated by the
canonical Lagrange anchor. In case $n=2$, we have a fourth-order
differential equation and a $2$-parameter set of the Lagrange
anchors generated by the canonical Lagrange anchor $V_{c}=1$ and the
operator of the second derivative $d^2/dt^2$. For the Pais-Uhlenbeck
oscillator this family is represented, in a different basis, by
equation (\ref{App2VPU}).

\section{Lagrange anchor for non-linear factorable systems}\label{AppD}

Here we derive the Lagrange anchor for equations (\ref{ODE2}) using
the formalism of Schwinger-Dyson constraints described in Appendix
\ref{AppA}.

The canonical Lagrange anchor for the Lagrangian theory (\ref{L})
gives the following SD constraints on the phase space of fields and
sources:
$$
\mathcal{P}\xi-U'(\alpha\xi-\beta\eta)-\frac{1}{\alpha}\bar\xi=0\,,\qquad
\mathcal{Q}\eta-U'(\alpha\xi-\beta\eta)+\frac{1}{\beta}\bar\eta=0\,,
$$
In the $\phi$-representation the corresponding SD constraints read
\begin{equation}\label{ODE2+V}
\mathcal{P}\mathcal{Q}\phi-\left(\frac1\alpha
\mathcal{Q}-\frac1\beta \mathcal{P}\right)\bar \phi-U'\left[(\alpha
\mathcal{Q}-\beta\mathcal{P})
\phi+\frac{(\alpha+\beta)^2}{\alpha\beta}\bar\phi\right]=0\,.
\end{equation}
Let us show that these constraints are in abelian involution. For
this end, we  make a linear canonical transformation from $(
\phi,\bar\phi)$ to the new variables
$$
\varphi=(\alpha\mathcal{Q}-\beta\mathcal{P})
\phi+\frac{(\alpha+\beta)^2}{\alpha\beta}\bar\phi\,,\qquad
\bar\varphi=-\mathcal{P}\mathcal{Q} \phi+\Big(\frac1\alpha
\mathcal{Q}-\frac1\beta \mathcal{P}\Big)\bar\phi\,.
$$
Since $\mathcal{P}$ and $\mathcal{Q}$ are Hermitian and commute, one
can easily find that
$$
\{\bar\varphi(x), \bar\varphi(x')\}=0\,,\qquad
\{\bar\varphi(x),\varphi(x')\}=\delta(x-x')\,,\qquad
\{\varphi(x),\varphi(x')\}=0\,.
$$
In terms of the new variables the SD constraints (\ref{ODE2+V}) take
the canonical form
$$
U'(\varphi)+\bar\varphi=0
$$
and the abelian involution is obvious. The inverse  canonical
$$
\phi=\Big(\frac1\alpha\mathcal{Q}-\frac1\beta\mathcal{P}\Big)\varphi-
\frac{(\alpha+\beta)^2}{\alpha\beta}\bar\varphi\,,\qquad
\bar\phi=\mathcal{P}\mathcal{Q}\varphi+(\alpha
\mathcal{Q}-\beta\mathcal{P})\bar\varphi\,.
$$
The SD constraint (\ref{ODE2+V}) involves the following Lagrange
anchor:
\begin{equation}\label{V}
V=\frac1\alpha\mathcal{Q}-\frac1\beta\mathcal{P}+
\frac{(\alpha+\beta)^2}{\alpha\beta}U''(\alpha\mathcal{Q}\phi-\beta\mathcal{P}\phi)\,,
\end{equation}
where the action of the matrix differential operator $U''$ is
defined by
$$
  U''(\varphi)\bar\varphi=\int dx\frac{\delta U'(\varphi)}{\delta\varphi(x)}\bar\varphi(x)\,.
$$
In the case $U=0$, the expression (\ref{V}) reduces to the Lagrange
anchor (\ref{VPQ}) that has been found in Appendix \ref{AppC}.

\end{document}